\documentclass[10pt,a4paper]{article}
\usepackage{a4wide}

\usepackage{color}
\usepackage{amsmath,graphicx}
\usepackage{amsfonts}
\usepackage{url}
\usepackage{amssymb}
\usepackage{amsmath}
\usepackage{caption}

\usepackage[labelformat=simple]{subcaption}

\usepackage{authblk} 
\usepackage{adjustbox}
\usepackage{multirow}

\usepackage{setspace}

\newtheorem{remark}{Remark}

\begin{document}
%
% paper title
% Titles are generally capitalized except for words such as a, an, and, as,
% at, but, by, for, in, nor, of, on, or, the, to and up, which are usually
% not capitalized unless they are the first or last word of the title.
% Linebreaks \\ can be used within to get better formatting as desired.
% Do not put math or special symbols in the title.
\title{RadioUNet: Fast Radio Map Estimation with Convolutional Neural Networks}

% author names and affiliations
% use a multiple column layout for up to three different
% affiliations

\author{\begin{center} Ron Levie$^{1,2}$, \c{C}a\u{g}kan Yapar$^3$\footnote{Equal contribution}, Gitta Kutyniok$^{1,4}$, Giuseppe Caire$^3$  \newline
\textcolor{white}{BLAN}
$^1$Department of Mathematics, LMU Munich \newline
\textcolor{white}{BLANK}
$^2$Institute of Mathematics, TU Berlin \newline
\textcolor{white}{BLA}
$^3$Institute of Telecommunication Systems, TU Berlin  \newline
$^4$Department of Physics and Technology, University of Troms{\o} \vspace{-4mm}
\end{center}}

\maketitle

% As a general rule, do not put math, special symbols or citations
% in the abstract
\begin{abstract}
In this paper we propose a highly efficient and very accurate deep learning method for estimating the propagation pathloss from a point $x$ (transmitter location) to any point $y$ on a planar domain. For applications such as user-cell site association and device-to-device link scheduling, an accurate knowledge of the pathloss function for all pairs of transmitter-receiver locations is very important. Commonly used statistical models approximate the pathloss as a decaying function of the distance between transmitter and receiver.  However, in realistic propagation environments characterized by the presence of buildings, street canyons, and objects at different heights, such radial-symmetric functions yield very misleading results. In this paper we show that properly designed and trained deep neural networks are able to learn how to estimate the pathloss function, given an urban environment, in a very accurate and computationally efficient manner. Our proposed method, termed RadioUNet, learns from a physical simulation dataset, and generates pathloss estimations that are very close to the simulations, but are much faster to compute for real-time applications. Moreover, we propose methods for transferring what was learned from simulations to real-life. Numerical results show that our method significantly outperforms  previously proposed methods. 
\end{abstract}

% no keywords
\textbf{Keywords:} 
Convolutional Neural Networks, Signal Strength Prediction, Radio Maps.

% For peer review papers, you can put extra information on the cover
% page as needed:
% \ifCLASSOPTIONpeerreview
% \begin{center} \bfseries EDICS Category: 3-BBND \end{center}
% \fi
%
% For peerreview papers, this IEEEtran command inserts a page break and
% creates the second title. It will be ignored for other modes.

\section{Introduction}
\label{sec:intro}

%\textcolor{green}{(Ron) change the deep learning notion ``channel'' to feature channel, to avoid confusion with wireless comm.}

In wireless communications, the pathloss is a quantity that measures the loss of signal strength (reduction in power, 
 or attenuation) between a transmitter (Tx) and receiver (Rx) due to large scale effects. The signal power attenuation may be caused by different factors, such as free-space propagation loss, 
 reflections and diffraction from buildings, waveguide effect in street canyons, and 
 obstacles blocking line of sight between Tx and Rx.
 The pathloss function (sometimes referred to as {\em path gain} function or {\em radio map}), 
 is a function that assigns to each Tx-Rx pair of locations $x, y$ the corresponding large-scale signal attenuation $G(x,y)$. Notice that in addition to the large scale effects, wireless propagation is also subject to small-scale fading, due to the superposition of scattered wavefronts with different phases at the Rx location. Such small-scale effects are typically modeled as a Gaussian random variable $H$ that, without loss of generality, can be normalized with unit second moment. Therefore, denoting by $Y = \sqrt{G(x,y)} H X + Z$ a signal sample at the Rx baseband output, 
 where $X$ is the transmitted signal sample with power $\textup{P}_{\textup{Tx}}$, 
 $H$ is the normalized small-scale fading, 
 and $Z$ is the additive noise with power spectral density $N_0$, 
 the received energy per sample is given by $\mathbb{E}[|Y|^2] = G(x,y) \textup{P}_{\textup{Tx}}/W + N_0$, where $W$ denotes the signal bandwidth, and the Signal to Noise Ratio (SNR) at the input of the Rx baseband processor is given by 
 ${\sf SNR} = \frac{G(x,y) \textup{P}_{\textup{Tx}}}{N_0W}$. In this paper we develop a deep learning method for estimating radio maps, which we call RadioUNet.

\subsection{Applications of Radio Maps}
Many applications in wireless communication explicitly rely on the knowledge of the pathloss function, and thus, estimating pathloss is a crucial task. 
 For example, in device-to-device (D2D) link scheduling, there exists a set of wireless devices that transmit signals to each other in pairs. A pair of devices that communicate defines a Tx-Rx link. The signal sent by a Tx is generally received by multiple Rxs beyond its intended destination, creating mutual interference between the links. 
 While the general information theoretic setting for this problem is the Gaussian interference channel, whose capacity region and optimal coding techniques are still an open problem in general,
 a huge amount of work has been devoted to the problem of scheduling subsets of links to be active on the same time slot and frequency subband, such that their mutual interference is sufficiently weak and the multiuser interference can be treated as Gaussian noise. 
 It turns out that in a particular regime of weak interference, {\em Treating Interference as Noise} (TIN) is information-theoretic approximately optimal \cite{TINJafar}. Furthermore, 
 efficient link scheduling and power control combined with TIN yields very good performance
in comparison with classical interference avoidance schemes such as CSMA \cite{bianchi2000performance}. 
A practical such link scheduling algorithm developed by Qualcomm is FlashLinQ \cite{wu2013flashlinq}.
Recent works on information-theoretic inspired D2D link scheduling include \cite{naderializadeh2014itlinq,TINCaire}, 
that significantly improve upon FlashLinQ. A recent more direct approach based on fractional programming optimization
is provided in \cite{shen2017fplinq}. All these schemes somehow assume that the
pathloss function between every Tx-Rx locations is known or can be accurately estimated via some probing scheme.
A deep learning approach to D2D link scheduling is proposed in \cite{SchYu}, which is implicitly based on the fact that interference is a decreasing function of distance and therefore that the pathloss function has a radial symmetry. Therefore, such scheme does not directly apply to more complicated urban propagation scenarios as considered in the present paper. 
From the above works it is clear that an accurate knowledge of the radio map for a specific environment is very important for efficient D2D links scheduling. 

Another classical use-case example of radio maps is base station assignment, or user-cell site association, where the goal is to assign a set of wireless devices to a set of cellular base stations. In order to decide which device to assign to which station, it is important to know the radio map (e.g., see \cite{userCellAssoc} and references therein). 

Some additional applications that rely on the knowledge of the pathloss function are fingerprint based localization \cite{FingerLoc}, physical-layer security \cite{PLS}, power control in multi-cell massive MIMO systems \cite{PowerCellular}, user pairing in MIMO-NOMA systems \cite{PowerNOMA}, precoding in multi-cell large scale antenna systems \cite{LSAS}, path planning \cite{pathUAV}, and activity detection \cite{activDet}.

%Another app: distributed antenna

\subsection{Radio Map Prediction}

A multitude of approaches for estimating the pathloss function have been proposed in the literature. For the sake of clarity, we can group these approaches in three categories. 

\emph{Data driven interpolation methods} assume that some measurements of the pathloss function are given at certain locations. These methods estimate the pathloss function at non-measured locations via some signal processing approach (e.g., {\em Kriging} \cite{stein2012interpolation}) and do not rely---or rely only lightly---on a model of the physical phenomenon. Beyond Kriging, other examples of such approaches are radial basis function interpolation \cite[Sect 5.1]{BishopNNPR}, tensor completion \cite{TensorCompletion}, support vector regression \cite{SVR}, and matrix completion \cite{MatComp}.

%\textcolor{black}{(Ron) Question: the radial basis function and tensor completion interpolation methods are simple interpolation %methods of points. Is the analysis of these methods based on some stochastic process? If not, perhaps we shouldn't call all %interpolation methods statistical methods. Maybe we can write ``These methods estimate the pathloss function at non-measured locations via some statistical \textbf{or} signal processing approach.''}%

%or tomography methods

\emph{Model-based data fitting methods} combine measurements of the pathloss function with a priori assumptions on the physical system to estimate the pathloss function at non-measured locations. For example, in tomography methods, the attenuation due to shadowing can be derived under some modeling assumptions from the so called spatial loss field (SLF), which in turn can be estimated from the measurements.
%obtained by a  bidimensional integral of the spatial loss field (SLF) of interest scaled by a weight function.  and estimating shadowing effects .  To this end, 
%
Here, various assumptions on the underlying SLF %and the weight function 
can be imposed, e.g., low-rank structure %with potential outliers of the SLF 
\cite{LowRankOutliers}, sparsity %of the SLF and the weight function 
\cite{BlindTomo}, and piecewise homogeneity \cite{AdaptiveBayesian,VariationalBayes}. % to name a few. 

%model-based simulations estimate the pathloss function based only on physical considerations, without any measurements.
Last, \emph{model-based prediction} estimates the pathloss function based only on available prior knowledge, e.g., physical considerations, without taking any measurements from the area of interest. Some examples are 
ray-tracing \cite{RayTracing},
dominant path model \cite{dominantPathUrban},
and empirical models, e.g. \cite{empV2V}.

\subsection{Radio Map Prediction Using Deep Learning}
\label{Radio Map Prediction Using Deep Learning}

Two recent papers proposed deep learning approaches for estimating radio maps  \cite{DocomoCNN,DocomoTwoStep}. There, the neural network is a function that returns an estimate of the pathloss for each input Tx-Rx locations. The network is trained on a fixed map and simulated pathloss values at a set of Tx-Rx locations. This procedure is a data-fitting method for the 4-dimensional (4D) function $G(x,y)$.\footnote{Notice that when $x$ and $y$ are points on the plane $\mathbb{R}^2$, the function $G(\cdot)$ has domain in $\mathbb{R}^4$.} Different city maps require re-training the network and each trained network describes a specific map. \textcolor{black}{In contrast, our RadioUNet learns to approximate the (outcome of the) underlying physical phenomenon, which is independent of a specific city map. Namely, the trained RadioUNet produces a radio map from any given Tx source and city map. We thus think of RadioUNet as a type of implicit simulation, given by the operations of its underlying convolution network.}
%In contrast, our RadioUNet learns the underling physical phenomenon, and executes a type of simulation, given by the operations of its underlying convolution network,
 %which interacts with any Tx source and city map. 
 Even when the map is fixed, we show that RadioUNet significantly outperforms previous deep learning proposed methods.
 
 \textcolor{black}{There are several more papers on pathloss prediction that use fully connected neural networks, which do not take the city map information into consideration, and use additional information such as the height of the transmitter/receiver or the distance between them. For example, see the survey \cite{PopoolaSurvey}, and the papers \cite{SOTIROUDISOptimal, SOTIROUDISDiff, PopescuUrban}. These methods are clearly unsuited to 
 predict the radio map as a function of the city map geometry, given as an input to the neural network, which is instead  the focus of this paper.}
 
Another recent work based on data-fitting to radio maps via deep learning, in the above fashion, is \cite{TLTilting}. The authors of \cite{TLTilting} also proposes a transfer learning approach to learn a radio map estimator corresponding to some antenna tilt $T_B$ from a radio map estimator of another tilt $T_A$. There, it is assumed that there is a large amount of data to train the tilt $T_A$, and a small amount of data for the tilt $T_B$. We also consider a transfer learning approach, in which we train a radio map estimator on a large dataset of simulations, and transfer it to real-life with the aid of a small dataset of real-life measurements.\footnote{Please see point 2) of Section \ref{subsec:contr} for the concept of ``real-life'' measurements used in this paper.}

\textcolor{black}{Slightly after our work, a convolutional autoencoder network was proposed for spectrum map interpolation \cite{RomeroICC}, where multiple transmitters with unknown locations operate simultaneously, and the city map is considered as an input along with measurements with known locations.}

% Another data fitting method:
% \textcolor{black}{A recent work \cite{TLTilting} proposes a transfer learning framework to estimate the Reference Signals Received Power (RSRP) radio maps corresponding to a different antenna tilt configuration by transferring the knowledge from another tilt configuration where no or only limited measurements from the target configuration is available. The authors provide a good overview of transfer learning including the categorizations for the questions like what/when/how to transfer?} \textcolor{black}{This can go in the intro} \textcolor{black}{(Ron) Write in short that they do something similar - freezing first layers and so on since they don't have much data. We also use the second UNet trained on DPM as initialization.}

\subsection{Our Contribution}
\label{subsec:contr}

In this paper we propose several versions of a radio map estimation method based on deep learning, which we term \emph{RadioUNet}. %\footnote{The source code of RadioUNet can be found at \url{https://github.com/RonLevie/RadioUNet}.}
 In our setting, we consider mobile devices/base stations in an urban environment. Our deep learning based methods are efficient, estimating the whole radio map within an area of $256^2 {\rm m}^2$ in an order of $10^{-3}$sec to $10^{-2}$sec, with root mean square accuracy of order $1{\rm dB}$, where the range of pathloss values from the noise floor to the maximal gain is 100dB. This is a mean accuracy of $1\%$ \textcolor{black}{(RMSE divided by the range)}.
Some preliminary results where reported in \cite{RadioUNetConf}, \textcolor{black}{were the proof of concept of estimating radio maps with UNets was illustrated on a preliminary dataset of simulations based only on the city maps (buildings) but not including details such as cars along the streets.}
\textcolor{black}{The source code of RadioUNet can be found at \url{https://github.com/RonLevie/RadioUNet}, and our dataset, RadioMapSeer, at \url{ https://RadioMapSeer.github.io}. For reproducibility, see the compute capsule at \url{https://codeocean.com/capsule/ea977fe8-d945-4a49-8326-0c687f96f8ff/tree}.}

\subsubsection{RadioUNet Methods}

Our radio map estimation methods are based on UNets \cite{UNet} and their compositions.
One version of RadioUNet (called RadioUNet$_{\rm C}$) only uses as input the city map (i.e., the geometry of the urban environment), 
the Tx location, and no pathloss measurements. % at specific points. 
Thus, this method can be categorized as model-based simulation. \textcolor{black}{However, as opposed to classical model-based simulation, our model is learned from training data. As such, on the one hand it does not have an explicit physically interpretable formulation, but on the other hand, its execution run-time (for the trained network) is much faster than existing model-based tools.} 
Another model that we propose (called RadioUNet$_{\rm S}$, with $\rm S$ for samples) takes  %as input the geometry of the urban environment, which may be also perturbed, the Tx location, and additional
as an additional input variable some measurements of the pathloss at a few locations. 
Thus, this method can be categorized as a model-based data fitting method. %Here, again, the model is implicit and learned from the data. 
Another optional input variable is the locations of cars along the streets, which help predicting the shadowing effect due to the penetration of the signal through cars.

\subsubsection{\textcolor{black}{The Training Data}}

We present a new dataset, called \emph{RadioMapSeer},\footnote{The dataset can be found at \url{ https://RadioMapSeer.github.io.}} of 56,000 simulated radio maps in different city locations and different Tx locations. Each simulation has a number of versions, generated using different types of {\em coarse} simulations (see details in Section \ref{General Setting}). 
In one type of simulation, cars are \textcolor{black}{generated} along the streets, \textcolor{black}{and affect the outcome of the simulation}. The cars serve as unpredictable obstacles perturbing the received signal strength. 
Alongside each simulation, the map of the city, the Tx locations and cars are also provided.

In addition, we present a smaller dataset of 1400 high accuracy simulations, with and without cars, called IRT4 (see Section \ref{General Setting}). 
In our setting, IRT4 serves as a surrogate for real-life measured radio maps, i.e., the effective ground truth with respect to which we calculate the prediction error. 
To imitate a realistic scenario, where the 1400 IRT4 simulations represent real-life measurements collected during a measurement campaign 
or even in real-time from user devices, each of the 1400 radio maps is only measured sparsely, e.g., we only have 300 receiver 
locations per map.
\textcolor{black}{At this point it is important to point out that the scope of our work is not to assess the accuracy of IRT4 
with respect to real-life measurements. The main idea here is that the coarse simulations and IRT4 share the same basic 
underlying propagation phenomenon, but IRT4 has additional finer details not present in the coarse simulations. 
One goal is then to develop methods to predict such fine details even though in training we have access to a (large) dataset of coarse simulations, but only to a (small) set of sparse measurements of IRT4. Of course, when RadioUNet is employed in practice, the refined phenomenon should be taken as the actual real-life measurements.}

\subsubsection{\textcolor{black}{Transferability to ``real-life''}}

%As discussed above, one important aspect that we address in this paper is how to transfer the RadioUNet, trained on coarse simulations, to real life. 
\textcolor{black}{One important aspect that we address in this paper is how to improve what RadioUNet learned from the coarse simulations to refined representations of the pathloss function. The ultimate goal for the sake of practical relevance is to transfer what RadioUNet learned from simulated data (the labeled training set) 
to real-life deployments. As a proof of concept, in the RadioMapSeer dataset we use the small set of high 
accuracy IRT4 simulations as a surrogate to actual real-life measurements.} 
Through this proof of concept, we show that the proposed methods learn the ``big-picture'' coarse phenomenon 
from the large coarse simulation datasets, and use the additional smaller 
set of IRT4 sparse samples to refine and adapt the RadioUNet to the refined phenomenon, 
using a small subset of the trainable parameters.
%
%We demonstrate \textcolor{experimentally} can learn to estimate the fine details of a more complex phenomenon. 
\textcolor{black}{Given the fact that ray tracing commercial tools are routinely used for wireless networks layout planning (e.g., see \cite{WinPropFEKO} or \url{https://www.remcom.com/wireless-insite-em-propagation-software}),  it is clear from engineering experience that 
high accuracy ray tracing can predict real-life measurements sufficiently well.  
This corroborates the significance of this proof of concept, where the role of real-life measurements (which are costly and difficult to obtain on such a large scale) is played here by the sparse but accurate IRT4 samples.} 

A second approach for transferability consists of training a RadioUNet to estimate radio maps 
from three {\em input feature channels}, the city map data, the Tx location, and some pathloss measurements. 
In this case the measurements are also taken from the same coarse simulation. \textcolor{black}{However, once trained, 
the RadioUNet is then applied in ``real life'', where real-life measurements of the pathloss are used as input. Again, in our experiments we test the transferability capability of this approach 
on the fine IRT4 simulation set.}

%To improve the transferability capability of the network in the above two scenarios, the network should give emphasis to the measurements, since they are taken as real-life pathloss values when the network is employed. The coarse simulated training data should give the network general guidelines to the big-picture behaviour of radio maps, but the network should not try to predict the fine details of the specific type of simulation, as these can be seen as artifact of the simulation. In other words, we want to make sure that RadioUNet ``understands'' that the simulations are not completely reliable, while samples are. 

%To produce unreliable simulations, we add a level of uncertainty to the training data. This is done by taking each simulation as a weighted average of DPM and IRT2 with random weights. The predictable part of these random simulations is exactly the phenomenon shared by DPM and IRT2, and the artifacts produced by DPM and IRT2 are unpredictable here. Since only predictable phenomena can be learned, RadioUNets based on unreliable simulations are ``clean'' radio map coarse simulators.

\subsubsection{Applications}

Our RadioUNet can be directly applied to any of the problems mentioned before, 
where an accurate knowledge of the pathloss function between any Tx-Rx pair of locations is useful. In a dynamic environment, the set of refined measurements can be provided in real-time from the mobile devices, along with their position. For the sake of space limitation, in this work we demonstrate the potential of our radio map estimation method with two \textcolor{black}{toy} applications.

\par{\noindent\textbf{i) Coverage classification.}}
We show how to predict the service area of a Tx, and conversely, show how to estimate the domain where the Tx creates  
small interference with other devices.
%In both of these problems the goal is to find the spatial area where the radio map is above or below some threshold. 

\par{\noindent\textbf{ii) Pathloss fingerprint based localization.}}
%We show how to use RadioUNet for pathloss fingerprint based localization.
Using the estimated radio maps of a set of devices/base-stations with known location, the location of some other device $d$ can be accurately computed if $d$ reports the received signal gains from the base-stations. %This is done by intersecting the level-sets corresponding to the reported gains of the radio maps of the different base-stations. Since RadioUNet gives good estimates of the ground-truth radio maps, a high quality localization can be computed by only knowing the map of the city and the locations of the base-stations.
%The pathloss based localization method can be potentially combined with additional fingerprint information.

%Where it can be used: (D2D, cell).

%We note that two recent papers proposed using deep learning methods for estimating radio maps or related notions.
%In \cite{DocomoCNN} the network receives as input a city map and two 2D images of distances of each pixel from the transmitter and receiver, along with additional system parameters. The network then outputs the path gain between the transmitter and the receiver. In \cite{DocomoTwoStep} spatial locations are classified as in the service zone or outside the service zone of a base station via a fully connected network.
%
%As opposed to these methods, which estimate pathloss between pairs of transmitter-receiver locations, our network computes the whole radio map at once. 

%Remark: in first DOCOMO, they train and test on the same single image? If so state that.

%In the journal paper implement then and compare using our units!
%Say why we do not compare agains their results: since error definition is weird.

%%%%%%%%%%%%%%%%%%%%%%%%%%%%%%%%%%%%%%%%%%%%%%%%%%%%%%%%%%%%%
\section{Background and Preliminaries}
\label{sec:background}

\subsection{Wireless Communication}
\label{wireless-comm}

%Path gain/pathloss/radio map definition. LSFC

Consider a general Gaussian interference network with $K$ Tx and $N$ Rx devices located over a certain region of the 2D plane. Following the {\em Generalized Degrees of Freedom} (GDoF) oriented model in \cite{TINJafar}, it is useful to normalize  the received signal such that the variance of the noise samples $N_0$ 
and the signal energy per symbol $\textup{P}_{\textup{Tx}}/W$ are both equal to 1, and define a parameter $P$ such that the normalized received signal at each $j$-th Rx is given by 
\begin{equation} \label{TIN-model}
Y_j = \sum_{i=1}^K \sqrt{P^{\alpha_{i,j}}} X_i + Z_j.
\end{equation}
where $\alpha_{i,j} = \frac{\log {\sf SNR}_{i,j}}{\log P}$ and ${\sf SNR}_{i,j}$ is the SNR between Tx $i$ and Rx $j$ as defined in Section \ref{sec:intro}. 
It turns out that the {\em GDoF region} of the underlying Gaussian interference network
(i.e., a high-SNR representation of the capacity region) is defined by the exponents $\alpha_{i,j}$. 
Furthermore, under certain conditions \textcolor{black}{of weak interference referred to as the TIN regime\footnote{In the information theoretic literature TIN stands for ``Treating Interference as Noise'' and the TIN regime is when TIN achieves the GDoF region, i.e., 
it is GDoF-optimal.} (see \cite{TINJafar,TINCaire} for the information-theoretic details) 
the GDoF region yields the actual {\em capacity region} within a bounded gap, independent of the SNR scale parameter $P$. 
These facts provide a strong evidence that the relevant notion of pathloss function is contained in the $\alpha$'s exponents, i.e., the pathloss function should be estimated in logarithmic scale (in dB).  
Furthermore, from the theory in \cite{TINJafar} it follows that 
negative values of the $\alpha$'s exponents are irrelevant, that is, for the GDoF region it is sufficient to take the positive 
part of the $\alpha_{i,j}$'s.} 
In practice, this means that we do not have to spend much effort in estimating very large negative values (in dB) 
of the pathloss function. As a matter of fact, it makes sense to truncate such function so that the received signal power 
is not too much smaller than the noise floor. 

Driven by the above considerations, we define the pathloss %function
 in dB scale 
as $\textup{P}_{\textup{L}} = (\textup{P}_{\textup{Rx}})_{\rm dB}-(\textup{P}_{\textup{Tx}})_{\rm dB}$,
where $\textup{P}_{\textup{Tx}}$ and $\textup{P}_{\textup{Rx}}$ denote the transmitted power and received power at the Tx and Rx locations, respectively. 
The truncation and rescaling of the pathloss function in dB scale in order to make it suitable for
the proposed deep learning estimation method is given in Sections \ref{sys-param} and \ref{Gray Level Conversion}.  

%%%%%%%%%%%%%%%%%%%%%%%%%%%%%%%%%%%%%%%%%%%%%%%%%%%%%%%%%%%%%%%%%%%%%%%
\subsection{\textcolor{black}{Deep Learning}}

%History..? was vaguely inspired by the brain..? Just a short intro to the section?

In this section we review some concepts from deep learning necessary for the understanding of this paper.
%\textcolor{black}{[GC: IT SEEMS THAT THIS SECTION HAS BEEN HEAVILY EDITED .. BUT I DON'T SEE ANY BLUE COLOR ... WE NEED TO EVIDENCE CHANGES IN BLUE SUCH THAT IT IS EASIER FOR THE REVIEWERS TO SEE THE CHANGES]} \textcolor{gray}{[RL: The problem is: how do you mark in color something that you deleted? I wrote in the response letter that when the title of a section is colored, it means that the whole section was revised. Otherwise I would need to color the whole section...]}

\subsubsection{Convolutional Neural Networks}

A \emph{Convolutional neural network} (CNN) is a popular deep learning architecture, typically used in machine learning applications in imaging science \cite{CNN_Nature,CNNLeCun}.
In our context,  a  \emph{feature map} is a function from a 2D grid to some $\mathbb{R}^N$, 
where $N$ is called the number of \emph{feature channels}. If $N=1$, we call the feature map a \emph{gray level image}.

A CNN is defined by aggregating the following five basic computational steps as the layers of the network. 
i) In a \emph{convolution layer} an input feature map is convolved with a filter kernel and added to some scalars called the \emph{bias}. The number of input feature channels and output feature channels need not coincide. More accurately, let $N$ be the number of input feature channels and $M$ the number of output feature channels. Let $f_1,\ldots,f_N$ be the components of the input feature map. Note that each $f_n$ is a gray level image, not a scalar. The components of the output feature map, $g_m$, are defined for every $m=1\ldots,M$ by
\begin{equation}
\label{eq:conv_layer}
    g_m = \sum_{n=1}^N f_n * y_{n,m} + b_m
\end{equation}
where $*$ denotes convolution,
and for each $m=1,\ldots,M$ and $n=1,\ldots,N$, $y_{n,m}$ is a gray level filter kernel, and $b_m$ is the $m$-th component of the bias.
ii) An \emph{activation function} is any function applied on the entries of a feature map, and a typical choice is ReLU, defined by $r(z)=\max\{0,z\}$. iii) A \emph{pooling layer} takes a feature map and down-samples it, e.g., by assigning the maximal entry of each $2 \times 2$ patch to the corresponding entry of the down-sampled feature map. iv) An  \emph{up-sampling layer} up-samples lower resolution feature maps to higher resolution ones. v) A \emph{fully connected layer} is a general linear operator/matrix applied on the feature map, and
added to some pre-defined bias. 
%
%A typical CNN architecture applies a sequence of convolution, activation, and pooling/up-sampling layers many times, optionally followed by some fully connected layers.  
%
A CNN architecture is defined by choosing how to combine the above layers, choosing the number of feature channels, and choosing the shapes of the filter kernels. The trainable parameters are the filters, the fully connected matrices, and the biases.

\subsubsection{UNets}

UNet is a special CNN architecture, introduced in \cite{UNet}, and used in a multitude of applications, including image segmentation \cite{FullyConv,Segnet,VNET,3DUnet}, video predicting \cite{UNetVideo}, super resolution/image inpainting \cite{UNetSupRes}, %Semantic scene understanding ?keep? [Semantic Foreground Inpainting from Weak Supervision?is this UNet?, Hallucinating Beyond Observation: Learning to Complete with Partial Observation and Unpaired Prior Knowledge], 
inverse problems in imaging \cite{InverseUNet}, image-to-image translation \cite{DualGAN}, and medical image analysis \cite{SurveyMedical} to name a few. 

UNets consist of convolution, pooling, up-sampling, and activation function layers, without fully connected layers.
The UNet architecture is divided into two paths. The first portion of the layers gradually contracts the image as the layers deepen, and gradually increases the number of feature channels. This path---also called the \emph{encoder}---is interpreted as a procedure for extracting ``concepts'' which become more complex/high-level and less spatially localized along the layers. The second portion of the layers---also called the \emph{decoder}---expands the image as the layers deepen and reduces the number of feature channels gradually. This path is interpreted as a procedure of combining/synthesizing the concepts, layer by layer, to lower-level concepts, and eventually to an output image. The decoder layers are derived by up-sampling lower resolution images, and thus lack high resolution information on their own. To provide high resolution information to the decoder layers, the feature maps  in the encoder layers are copied and concatenated to the corresponding feature maps of the decoder layers having the same resolution. This copying between non-neighboring layers is called \emph{skip connection}.  %UNet architecture is represented in Table **.

%We write down UNets explicitly as follows. Consider a UNet $U$ based on $L$ layers. Let $\mathbf{p}_l$ denote the vector of all learnable parameters of layer $l$ of the UNet. Namely, $\mathbf{p}_l$ is a list that concatenates all of the entries of the different filters and the different biases of Layer $l$. For any $l=1,\ldots,L$, denote by $U^l$ the function that maps the feature map of layer $l$ to the feature map of layer $l+1$. Namely, $U^l$ applies a convolution plus bias step of the form (\ref{eq:conv_layer}), followed by an activation function and optionally pooling or up-sampling. To emphasize the reliance of $U^l$ on $\mathbf{p}_l$ we denote $U^l_{\mathbf{p}_l}$, and $U^l_{\mathbf{p}_l}$ applied on the feature map $\mathbf{f}$ of layer $l$ is denoted by $U^l_{\mathbf{p}_l}(\mathbf{f})$.

Let $\mathbf{p}$ denote the concatenation of all learnable parameters of the UNet, and let $U_{\mathbf{p}}$ denote the UNet, mapping input images $\mathbf{f}$ to output images $U_{\mathbf{p}}(\mathbf{f})$.
%\[U_{\mathbf{p}} = U^1_{\mathbf{p}_1} \circ U^2_{\mathbf{p}_2} \circ \ldots \circ U^L_{\mathbf{p}_L}.\]
 %The output of the UNet on the input signal $\mathbf{f}$ is given by 
%\begin{equation}
%\label{eq:comp_UNet}
%    U_{\mathbf{p}}(\mathbf{f}) = U^L_{\mathbf{p}_L}\Big( U^{L-1}_{\mathbf{p}_{L-1}}\big(\ldots U^1_{\mathbf{p}_1}(\mathbf{f}) \ldots \big)\Big).
%\end{equation}
%
In supervised learning, a \emph{training set} of many example inputs images  $\mathbf{f}_k$ and corresponding desired output images $\mathbf{g}_k$ are given, where $k=1,\ldots,K$ and $K$ is the size of the dataset. The goal is to fine tune the parameters $\mathbf{p}$ of $U_{\mathbf{p}}$ so that $U_{\mathbf{p}}(\mathbf{f}_k) \approx \mathbf{g}_k$ for every $k=1,\ldots,K$. 
This is typically done by some variant of gradient descent, and the loss function to be optimized is typically of the form 
\begin{equation} 
\mathcal{L}(\mathbf{p})=\frac{1}{K}\sum_{k}\left\| \mathbf{g}_k - U_{\mathbf{p}}(\mathbf{f}_k) \right\|
\end{equation}
for some norm, e.g., the root mean square norm. In \emph{stochastic gradient descent} (SGD), $k$ runs over one batch in each iteration. 

\subsubsection{Curriculum Learning}

The SGD optimization procedure (and its variants) explores configurations of the parameters only along the 1D path of descent, which might miss good configurations.  Namely, SGD searches the parameter space in a highly non-exhaustive manner. 
Thus, the expressive capacity of a network does not guarantee high quality trained networks. %This happens since in gradient descent the network explores only configurations of the parameters along one path, that might miss good configurations. %
It is thus often important to lead gradient descent in a more deliberate way, and in some sense to ``micro manage'' the exploration of parameter configurations in the optimization process. One approach for achieving this is called \emph{curriculum learning} \cite{Bengio_cur}. In curriculum learning, training is divided into a \emph{curriculum}, namely, a list of optimization problems, where the optimal solution of the previous problem is used as the initial guess for the next optimization problem. The idea is to first teach the network how to solve an easy to learn simplified version of the problem, and gradually to increase the complexity of the problem until reaching the original formulation of the loss function.

\subsubsection{Transfer Learning}

In some learning scenarios the training data does not represent exactly the data in the target application (e.g., when a large enough training set is difficult or costly to obtain).
It is thus important to know whether the network, trained on one data distribution, performs well for another data distribution. 
The idea of training in one domain and testing in another domain is called \emph{transfer learning} \cite{transfer1,transfer2}. 
The capacity of a network to perform well in new domains is called its \emph{transferability}.

\section{The RadioMapSeer Dataset}

In this section we introduce RadioMapSeer, a dataset of city maps with corresponding simulated radio maps that we have created and made available for this work.

\subsection{General Setting}
\label{General Setting}

\begin{figure}[!ht]
   \centering
   \begin{subfigure}{0.8\textwidth}\centering
   \subfloat[][DPM]{\includegraphics[width=.24\textwidth]{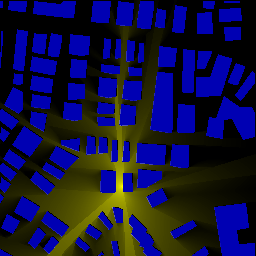}}\ %\quad
   \subfloat[][DPM with cars]{\includegraphics[width=.24\textwidth]{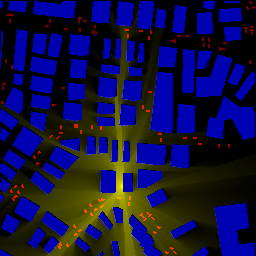}}\ %\quad
   %\subfloat[][IRT2]{\includegraphics[width=.24\textwidth]{Thr2FusedIRT2.png}}\quad
   \subfloat[][IRT4]{\includegraphics[width=.24\textwidth]{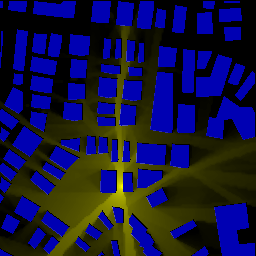}}\ %\quad%\\
   \subfloat[][IRT4 with cars]{\includegraphics[width=.24\textwidth]{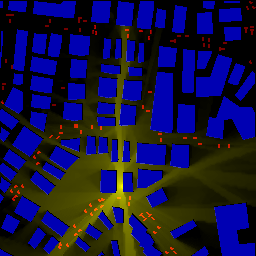}}
   %\subfloat[][DPM with cars]{\includegraphics[width=.3\textwidth]{Thr2FusedDPMcars.png}}\quad
   %\subfloat[][IRT2 with cars]{\includegraphics[width=.3\textwidth]{Thr2FusedIRT2cars.png}}\quad
   %\subfloat[][IRT4 with cars]{\includegraphics[width=.3\textwidth]{Thr2FusedIRT4cars.png}}\quad
   %\caption*{Simulated models }
   \label{fig:models}
   \end{subfigure}
   \begin{subfigure}{0.19\textwidth}\centering \vspace{2mm}
  \includegraphics[width=0.96\linewidth]{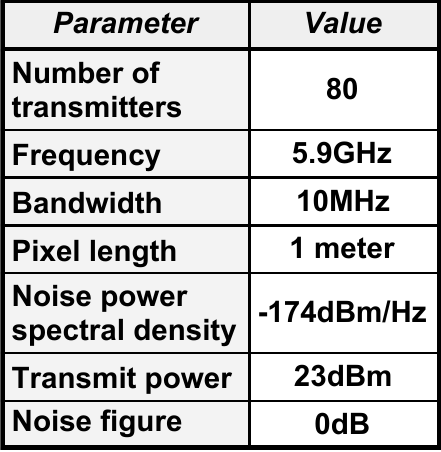}
  \caption{RadioMapSeer Dataset parameters}
	\label{table:RMSpars}
   \end{subfigure}
   \caption{\small RadioMapSeer examples and parameters. Buildings are blue, cars red, and pathloss yellow.}% {\color{red} GC: it seems that this figure is never referenced in the paper ... you have to reference the figure in the text of the paper otherwise it is unclear what this is ...} \textcolor{green}{(Ron) Get rid of this for the journal, and write the system parameters in the text.}}
   \label{RadioMapSeerExmp}
\end{figure}

The RadioMapSeer dataset consists of 700 maps, 80 transmitter locations per map, and corresponding coarsely simulated radio maps. \textcolor{black}{The coarse simulations are generated using 
the Dominant Path Model (DPM)  method \cite{dominantPathUrban}
and Intelligent Ray Tracing (IRT) \cite{IRT} based on 2 interactions of the rays with the geometry, referred to here as IRT2.
In addition, we have also generated fine simulations using IRT with 4 interactions (IRT4), for the first two transmitters of each map.}
The city maps are taken from \emph{OpenStreetMap} \cite{OpenStreetMap} in the cities Ankara, Berlin, Glasgow, Ljubljana, London, and Tel Aviv. 
\textcolor{black}{We set the heights of the transmitters, receivers and buildings as 1.5m, 1.5m, and 25m, respectively,} \textcolor{black}{which is relevant to  device-to-device 
scenarios (see Section \ref{sys-param} for more details). All simulations were computed using the software \emph{WinProp} \cite{WinPropFEKO}.
For different situations (e.g., campus networks, cellular networks) one should generate new data sets 
accordingly.} Some example radio maps from the dataset are shown in Figure \ref{RadioMapSeerExmp}.
All simulations are saved as dense sampling of the radio map in a 2D grid of 256$\times$256 m$^2$.

\subsubsection{Maps and Transmitters}

Each map is $256\times256m^2$ where buildings and roads are saved in the dataset as polygons. 
Each map is also converted to morphological 2D image (binary \textcolor{black}{$0/1$ pixel values with no intermediate gray levels}) of $256 \times 256$ pixels, where each pixel represents one square meter. 
The interior of the buildings \textcolor{black}{has pixel value $=1$}, and the exterior of the buildings  \textcolor{black}{has pixel value $=0$}. The transmitter locations are stored as numerical 2D values, and also given as morphological images, where the pixel in which the transmitter is located \textcolor{black}{has value $1$} and the rest is $0$. 
Along with the city maps, roads are given both as polygonal lines and as morphological images with 1 on the road and zero outside. Cars are generated along and aside roads, and given as separate morphological images.% Another version of each simulation,  with the effect of the cars, is also given.

\subsubsection{Coarsely Simulated Radio Maps}

The coarse radio maps were generated using DPM and IRT2 with the radio network planning software \emph{WinProp} \cite{WinPropFEKO}. 
Each simulated radio map stores at each pixel the pathloss between the pixel location and the transmitter location in dB.

To represent uncertainty in the dataset we consider two cases. 
First, a set of simulations on all city maps including the cars is produced using DPM and IRT2. 
These simulations are perturbations of the simulations based on the city map alone, \textcolor{black}{without cars}. 
We moreover provide separate datasets of perturbed city maps, where in each map of the original dataset $m$ buildings are missing. 
We provide four such datasets with $m=1,\ldots,4$. 

\subsubsection{Higher Accuracy Simulations}

An additional smaller dataset of higher accuracy simulations is provided by using IRT4 with the same \emph{WinProp} radio network planning software.  
Here, for each of the 700 maps we consider two transmitter locations. The goal of the higher accuracy simulations is to provide means of testing whether the network, trained on simulations, 
performs well in \textcolor{black}{refined representations of pathloss functions. As already said before, we reiterate here that the high accuracy simulation serves as a surrogate for the real-life physical phenomenon, and it is useful for demonstrating
the transferability property of our scheme. }

\subsubsection{Pathloss Scale}

The pathloss values $P_{\textup{L}}$ are converted to \emph{gray level} pixel values between 0 and 1
(see Subsection \ref{Gray Level Conversion}).  \textcolor{black}{Therefore, each radio map is a gray level image of size $256 \times 256$.}

\subsection{System Parameters} 
\label{sys-param}

This study was originally motivated by device-to-device communications for safety in the context of 
 intelligent transportation systems (ITS), currently based on the IEEE 802.11p standard. 
Accordingly, we consider a signal bandwidth $W$  of 10MHz in the 5.9GHz band. We choose the transmitter power and thermal noise power spectral density as $(\textup{P}_{\textup{Tx}})_{\rm dB} = 23$dBm and $(N_0)_{\rm dB} =-174$dBm/Hz in compliance with IEEE 802.11p and assume an idealistic noise figure of 0dB at receivers (cf. Table \ref{table:RMSpars} for a summary of the system parameters).

We express by $(\mathcal{N})_{\rm dB} =  10\log_{10} W N_0 + \textup{NF}$ the noise floor in dB, with NF being the noise figure.
%\textcolor{black}{with or without $10\log_{10} 10^6$?? It's conversion from MHz. See \url{https://en.wikipedia.org/wiki/Minimum_detectable_signal} for definitions of noise figure, floor Actually setting SNR=0 is called the noise-floor.}.
We consider the points  where the received signal power $(\textup{P}_{\textup{Rx}})_{\rm dB} = \textup{P}_{\textup{L}}+ (\textup{P}_{\textup{Tx}})_{\rm dB}$  yields a signal-to-noise ratio above a desired SNR level, i.e. the points where $(\textup{SNR})_{\rm dB} = (\textup{P}_{\textup{Rx}})_{\rm dB} - (\mathcal{N})_{\rm dB} \geq \textup{SNR}_{\textup{thr}}$ holds.  Solving this for $\textup{P}_{\textup{L}}$ we get the threshold $\textup{P}_{\textup{L,thr}}$ for the pathloss
\begin{equation}\label{eq:thr}
    \textup{P}_{\textup{L}}\geq \textup{P}_{\textup{L,thr}} = -(\textup{P}_{\textup{Tx}})_{\rm dB} + \textup{SNR}_{\textup{thr}} + (\mathcal{N})_{\rm dB}.
\end{equation}
We call $\textup{P}_{\textup{L,thr}}$ the \emph{pathloss threshold}.
Consider for example the SNR requirement that the received signal power should be above the noise floor, i.e., when $\textup{SNR}_{\textup{thr}}=0$. With the choice of parameters in Table \ref{table:RMSpars}, we find $\textup{P}_{\textup{L,thr}} = -127$dB. 
%\textcolor{black}{Is there a name for $P_{\textup{L,thr}}$? I would like to call is by a name, something like a version of the noise floor...} \textcolor{black}{I referred to it as the pathloss threshold for a desired minimum SNR at receiver, without using any special name.}

One task of RadioUNet is to extract the area in the city map above 
%the desired signal-to-noise ratio $\textup{SNR}_{\textup{thr}}$,
the noise floor, 
given an input city map and transmitter location. To do this, the network must learn the physical phenomenon both above and below the noise floor.  We thus truncate the pathloss values below another threshold $\textup{P}_{\textup{L,trnc}} < \textup{P}_{\textup{L,thr}}$. We choose $\textup{P}_{\textup{L,trnc}}$ such that the difference between the maximum pathloss $M_1$ in the dataset and $\textup{P}_{\textup{L,thr}}$ is approximately four times greater than the difference between $\textup{P}_{\textup{L,thr}}$ and $\textup{P}_{\textup{L,trnc}}$, i.e., $M_1-\textup{P}_{\textup{L,thr}} = 4(\textup{P}_{\textup{L,thr}}-\textup{P}_{\textup{L,trnc}})$. The maximum and the minimum pathloss in the dataset are -47.84dB 
%\textcolor{black}{(If the transmit power is positive, and pathloss at the transmitter is approximately zero, how come you get something negative? - reason - average over a pixel - explain this)} 
and -186.41dB, respectively. Note that the maximum is -47.84dB and not 0dB since the pathloss is integrated over 1m$^2$ pixels.  To meet the previously mentioned condition, we set $\textup{P}_{\textup{L,trnc}} = -147$dB. Since any signal below $\textup{P}_{\textup{L,thr}}$ cannot be detected in practice, and is only used in simulation for theoretical reasons, we call $\textup{P}_{\textup{L,trnc}}$ the \emph{analytic noise floor}.
%
%Note that by \eqref{eq:thr} we have $\textup{P}_{\textup{L,thr}} = -\textup{P}_{\textup{Tx}} + \textup{SNR}_{\textup{thr}}  + \textup{NF} + 10\log_{10} W  N_0$. Hence, any choice of the parameters on the RHS that results in the same pathloss threshold $\textup{P}_{\textup{L,thr}}$ has the same radio map.

 \subsection{Gray Level Conversion}
 \label{Gray Level Conversion}

We convert the pathloss values $\textup{P}_{\textup{L}}$ to pixel values between 0 and 1 as follows.
Denote by $M_1$ the maximal pathloss in all radio maps in the dataset, and define  $f=\max\{\frac{\textup{P}_{\textup{L}}-\textup{P}_{\textup{L,trnc}}}{M_1-\textup{P}_{\textup{L,trnc}}},0\}$. Here, $f=0$ represents anything below the analytic noise floor, and $f=1$ represents the maximal gain at the transmitter.
Any intermediate value is referred to as a \emph{gray level}.

Let us explain the importance of our gray level conversion when evaluating the performance of any pathloss estimation.
We evaluate performance of any approximation $\tilde{f}:\mathcal{D}\rightarrow\mathbb{R}$ of a signal/image $f:\mathcal{D}\rightarrow\mathbb{R}$, where $\mathcal{D} = \{x_n\}_n$ is some finite grid in $\mathbb{R}^2$, via the normalized mean square error (NMSE)
\begin{equation}
    E= \frac{\sum_n |\tilde{f}(x_n)-f(x_n)|^2}{\sum_n |f(x_n)|^2}.
    \label{eq:10}
\end{equation}
The numerator in (\ref{eq:10}) represents the absolute error, and the denominator represents the global magnitude of $f$.
The coefficients $|\tilde{f}(x_n)-f(x_n)|^2$ and $|f(x_n)|^2$ having larger values affect the outcome of $E$ the most, and small values are negligible. It is thus crucial to express the signal $f$ in a representation in which the important parts of the signal obtain large values.

In our case, the representation of the radio map should be constructed in such a way that small powers contribute small values to $E$. Indeed, locations of small power represent a weak signal.
If we represent the radio map as standard pathloss, in dB, the smaller the power in a certain location, the higher the magnitude of the pathloss, with negative sign. When the power goes to zero, the pathloss diverges to $-\infty$. In this representation, locations of a weak signal dominate the global magnitude of the radio map, and in general define a misleading concept of the ``size'' of the radio map. A similar situation occurs for the absolute error (the numerator of (\ref{eq:10})). 

As discussed in Section \ref{wireless-comm}, motivated by the GDoF region of a Gaussian interference network,
we know that very large negative values of the pathloss are effectively irrelevant and should not dominate the 
overall error.  Our gray level conversion resolves this issue. Indeed, anything below the noise floor, or more generally, below $\textup{P}_{\textup{L,trnc}}$, is deemed ``too small to be interesting'', and set to zero. 
In contrast, the values of higher power, which are most important, are transformed to levels close to $1$. 
We note that papers like \cite{TensorCompletion,MatComp,LowRankOutliers} suffer from the aforementioned shortcoming, and it is thus difficult to interpret their reported performance. 

When root mean square error (RMSE) is used, the gray level error is simply a scaling of the RMSE of the pathloss in dB (up to the truncation below the analytic noise floor). More precisely, we have
\[\sqrt{\sum_n |\tilde{\textup{P}}_{\textup{L}}(x_n)-\textup{P}_{\textup{L}}(x_n)|^2}=C\sqrt{\sum_n |\tilde{f}(x_n)-f(x_n)|^2},\]
where $\textup{P}_{\textup{L}}$ is the pathloss in dB. For $\textup{SNR}_{\textup{thr}}=0$ we have $C=80$.

\begin{remark}
\textcolor{black}{It is worthwhile noticing that the RMSE of the pathloss in dB scale (up to rescaling as explained above) comes here not by accident, and in fact it is a very sensible choice for the approximation error from a communication theory significance viewpoint. 
The pathloss $\textup{P}_{\textup{L}}$ in natural scale operates as a multiplier of quantity $\frac{\textup{P}_{\textup{Tx}}}{N_0W}$ in order to yield the SNR at the Rx location.
If $\textup{P}_{\textup{L}}$ is perfectly known, then the pathloss prediction result can be immediately translated into a receiver rate result, using either the Shannon capacity formula for the appropriate channel model, 
or a table of rates versus Rx SNR according to the family of coding and modulation schemes 
specified by a particular communication standard. 
On the other hand, if $\textup{P}_{\textup{L}}$ is known up to the approximation RMSE $\sigma$ in dB scale, we can model such approximation error as normal with standard deviation $\sigma$.\footnote{Notice that approximating a random variable with zero mean and given standard deviation $\sigma$ as normal $\mathcal{N}(0,\sigma^2)$ yields the maximum-entropy approximation, which has been widely used in statistics \cite{cover1999elements}.}
This means that the pathloss $\textup{P}_{\textup{L}}$ in linear scale is known up to a log-normal shadowing fluctuation 
with parameter $\sigma$. Then, one can easily provide performance guarantees in terms of rate 
versus outage probability, where the latter can be evaluated from the tail of the 
log-normal distribution. Since these results are highly dependent on the system assumptions, 
we provide here just a simple example, leaving to the reader the generalization to to any system
of choice. Using the capacity formula for the AWGN channel and assuming a Tx transmission rate 
$R$ bit/s/Hz, a receiver with pathloss (in linear scale) $\textup{P}_{\textup{L}} \times \Delta$, 
where $\textup{P}_{\textup{L}}$ is the exact value and $\Delta$ is the log-normal error, 
can decode successfully if $R$ is strictly less than  
the capacity $\log_2 \left ( 1 + \textup{P}_{\textup{L}} \times \Delta \times \frac{\textup{P}_{\textup{Tx}}}{N_0W} \right )$. 
Then, the probability of (block) decoding error for such ideal 
coded system is given by the {\em information outage probability}, i.e., the probability that the 
Shannon capacity falls below the transmission rate $R$ \cite{biglieri1998fading}. 
Using the fact that $10 \log_{10} \Delta$ is normally distributed $\mathcal{N}(0, \sigma^2)$, we obtain immediately: \[ P_{\rm out}(R) = 1 - Q \left ( \frac{(2^R-1)_{\rm dB} + (N_0 W)_{\rm dB} - (\textup{P}_{\textup{Tx}})_{\rm dB} - (\textup{P}_{\textup{L}})_{\rm dB}}{\sigma} \right ),  \]
where $Q(x) = \int_x^\infty \frac{1}{\sqrt{2\pi}} e^{-u^2/2} du$ is the Gaussian tail function. 
Hence, knowing $\textup{P}_{\textup{L}}$ and the RMSE $\sigma$ it is possible to obtain 
results in terms of receiver rate $R$ for a given guaranteed 
block error probability $P_{\rm out}(R)$ at any location $x_n$ of the map. 
\hfill $\lozenge$}
\end{remark}

\section{Estimating Radio Maps via RadioUNets}

In this section we introduce a number of methods,  collectively called RadioUNet, that learn to estimate radio maps in different scenarios. We evaluate the accuracy of the proposed methods and compare them to state-of-the-art.

\subsection{Motivation for RadioUNet}

\textcolor{black}{UNets have been extensively applied to imaging problems in the past few years with resounding success, and are considered to be a baseline method for image-to-image tasks \cite{UNetBase}. Our problem can be seen as mapping an image representing the city and Tx to an image representing the radio map, and hence using UNets is a natural choice.
One advantage of using UNets in our case is that they respect the translation invariance symmetry of the physical phenomenon. Namely, this symmetry is built in to RadioUNet, and requires no training. Another strong point of UNets is the encoder-decoder interpretation, as we discuss next.}

\begin{figure}
	\centering
	\begin{subfigure}{0.48\textwidth}\centering
	\begin{subfigure}{0.48\textwidth}\centering
		\includegraphics[width=1\linewidth]{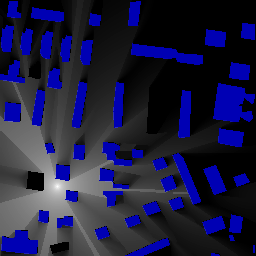}
	\end{subfigure}%
	\hfill
	\begin{subfigure}{0.48\textwidth}\centering
		\includegraphics[width=1\linewidth]{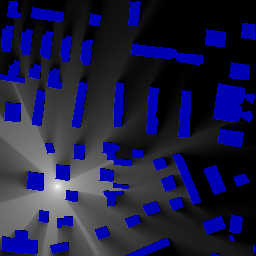}
	\end{subfigure}\hfill \newline
	\vspace{2mm}
	
	\begin{subfigure}{0.48\textwidth}\centering
		\includegraphics[width=1\linewidth]{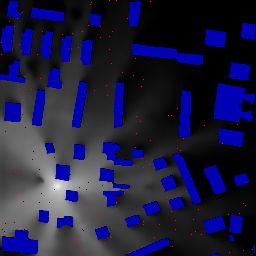}
	\end{subfigure}%
	\hfill
	\begin{subfigure}{0.48\textwidth}\centering
		\includegraphics[width=1\linewidth]{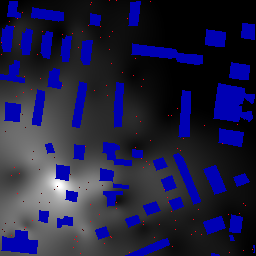}
	\end{subfigure}%
	\caption{ Comparison of RadioUNet with RBF with four missing buildings in the input. From left to right. 1:Ground truth radio map. 2:RadioUNet$_{\rm c}$ with all buildings. 3:RadioUNet$_{\rm S}$ with missing buildings. 4:RBF. The measured 127 locations for both RadioUNet$\rm S$ and RBF are marked in red. For RBF the transmitter is also a measurement, and the known buildings are set to zero post-processing. Known buildings are marked in blue.} 
	\label{fig:First_figure}
	\end{subfigure}
	\begin{subfigure}{0.5\textwidth}
	\centering
    \includegraphics[width=0.9\linewidth]{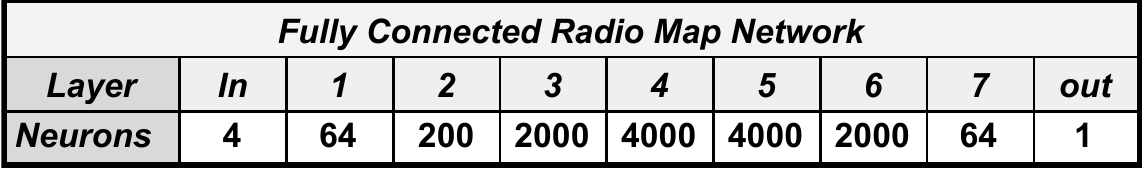}
		\caption{\small Architecture of the MLP of \cite{DocomoTwoStep}.}
		\label{fig:Second05_figure}
    	\vspace{5mm}
    \includegraphics[width=0.95\linewidth]{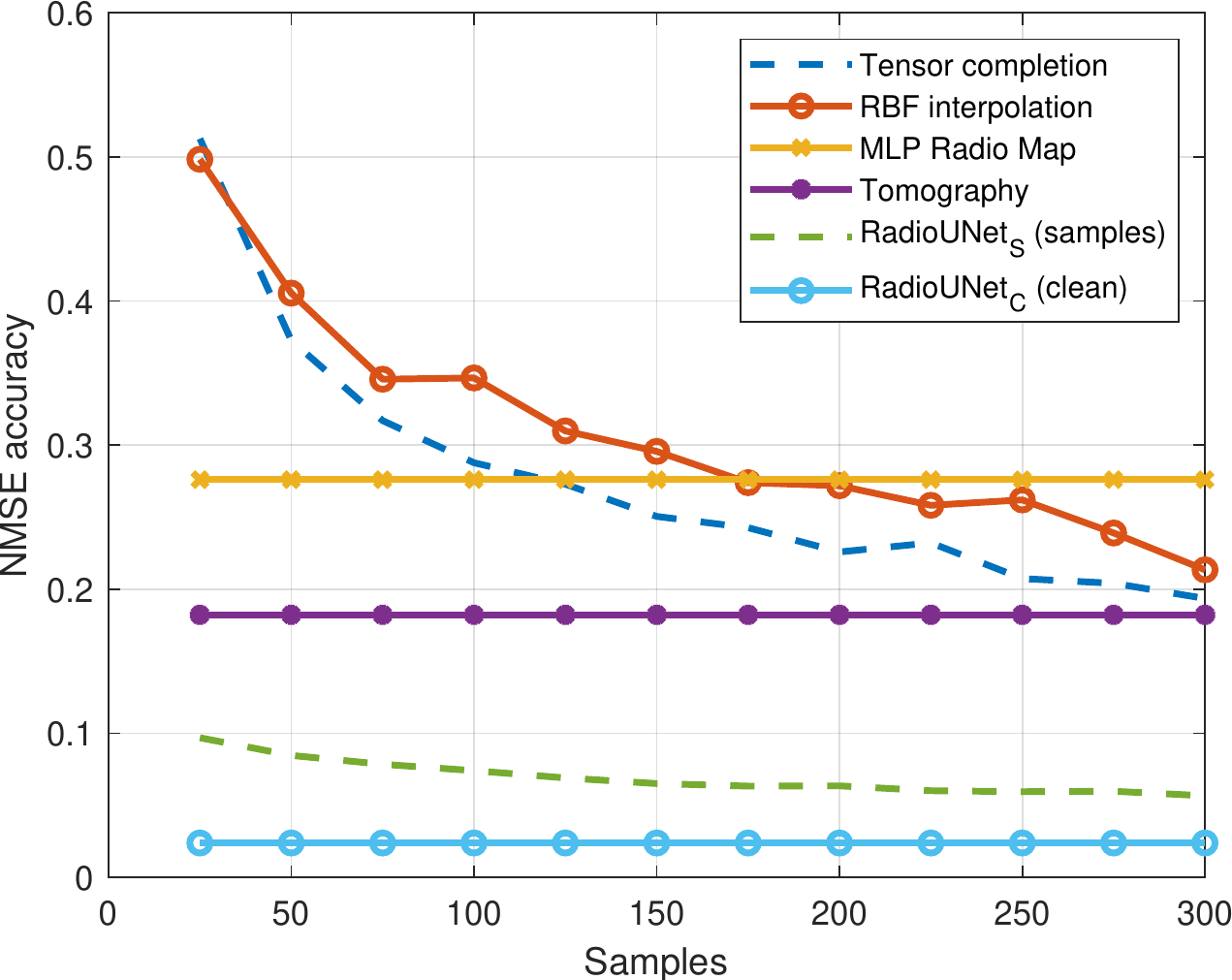}
    \caption{ Estimation error of the radio map reconstruction methods as a function of the number of measurements. We chose Map 12 from the test set, on which RadioUNet performs worse than the average test map (three times the average NMSE). RadioUNet$_{\rm C}$, Tomography, and deep learning one-step (MLP), are based on no samples, and are given as horizontal baselines. }
    \label{fig:Second_figure}
	\end{subfigure}
	\caption{Comparison or RadioUNet with state-of-the-art.}
	\label{fig:50}
\end{figure}

In Fig.~\ref{fig:First_figure} we show an example of a ground truth radio map generated by simulation, and the estimated radio map computed by the RadioUNet$_{\rm C}$ and RadioUNet$_{\rm S}$. 
%\textcolor{black}{It seems that at this point you haven't yet defined RadioUNet$_{\rm C}$ and RadioUNet$_{\rm S}$, so this statement comes unexpected and may confuse a reviewer ...if these are introduced later, say here (see Section [blabla])}. 
%We also show the estimation by RBF interpolation.
Aside from the low quantitative error, RadioUNet seems to synthesize radio maps from the urban geometry which qualitatively captures the correct shadow patterns. Note that the results in Fig.~\ref{fig:First_figure} are representative of the general quality of RadioUNet.
One might naively interpret the success of the RadioUNet by postulating that it learns to mimic a physical model, like ray-tracing or some differential equation like Maxwell's equations. However, we believe that this is a misleading viewpoint. A more reasonable interpretation follows from the encoder-decoder description of general UNets. In the encoder path, the RadioUNet extracts complicated concepts about the geometry of the urban environment and the mutual relationship between the different geometric features, their location, and the location of the transmitter. Then, in the decoder path, the RadioUNet uses these concepts to synthesize the radio map. Thus, RadioUNet is based on extracting and analyzing \emph{global} information about the urban environment, as opposed to classical physical models that are based on \emph{local} information, like collisions with the geometry in ray-tracing and derivatives in differential equations. In this viewpoint, it is more fitting to compare RadioUNet to a highly skilled artist that draws radio maps from his/her perception of the urban environment as a whole, rather than comparing to a classical local physical model.

\subsection{Different Setting in Radio Map Estimation}

We consider the following scenarios for the input of the UNet, the map of the city, the learning setting, and the properties of the simulated dataset. The problem setting can be any combination of the choices presented in Subsections \ref{Network input scenarios} and \ref{Learning scenarios}.

\subsubsection{Network Input Scenarios}
\label{Network input scenarios} $ $

\par{\noindent\textbf{City map and transmitter location.}}
In the first case, the UNet receives as input the map of the city and the Tx location as morphological images. %The Tx feature channel 
%is an image where the pixel in which the Tx is located \textcolor{black}{has value 1}, and the rest is \textcolor{black}{0}. 
From these two input feature channels the network estimates the radio map.

In this {\em accurate map scenario}, if the simulated dataset without cars is used, then the map without cars is given as input, while if the simulated dataset includes cars, then the map without cars is given as one feature channel, and the cars in an additional input feature channel.

When the map is accurate and the simulated data used for training is assumed to represent reality accurately, 
the radio map is uniquely determined by the map and the Tx location. 
Thus, the input feature channels are sufficient for high quality radio map reconstruction.
\vspace{1mm}

\par{\noindent\textbf{City map, transmitter location, and measurements.}}
In the second case, the UNet receives as input the two/three feature channels of map and Tx location as before, and an additional feature channel of measurements of the ``true'' radio map. The measurements are taken at some locations on the true map, i.e., their values are sampled from the target ``ground truth''. This third feature channel is given as a gray level image, where in the pixels corresponding to the locations of the measurements the gray level value is the measurement. Non-measured pixels are set to zero. The network \textcolor{black}{estimates} the radio map from these three/four input feature channels.

This scenario is useful when the ``nominal'' map given as input feature channel does not represent reality completely accurately. Hence, the network learns a hybrid of a radio map estimation method based on the given map, which is not completely reliable, and an interpolation method of the accurate pathloss measurements.
In this {\em non-accurate maps scenario}, a perturbed version of the ground truth maps is given as input to the UNet. 
We consider two types of perturbations: 1) the map is given with a one to four missing buildings; 2)  the map is given without cars, but the ground truth simulation is computed with the cars. 

Another source of inaccuracy, for which relying on measurements is useful, is the fact that training is done against coarse simulations, which are only approximations of reality (or, in our setting, approximations of IRT4).

\subsubsection{Learning Scenarios}
\label{Learning scenarios} $ $

\par{\noindent\textbf{Large and dense simulation dataset.}} Here, the network is trained in supervised learning to predict a large dataset of 2D gray-level images representing dense measurements of radio maps on a fine grid. The images are the DPM simulations, the IRT2 simulations, both with or without cars, or random combinations of DPM and IRT2. 
In particular, the goal in the randomized simulation is to push the network to learn that it can only rely on the simulations for the big-picture behavior of radio maps, shared both by DPM and IRT2, but not on the fine details. This pushes the network to use additional information for refining the estimations, like the input measurements if given, or the smaller dataset of sparse IRT4 if given.

Transferring the trained network to the ground truth (IRT4 or real-life maps) 
is a \emph{zero-shot generalization}. Namely, the network only learned to estimate the coarse simulations, not ever seeing the ground truth phenomenon, and we rely on the accuracy of the simulations, and optionally on the measurements, to predict the ground truth radio maps. 

\textcolor{black}{In case measurements are given as an input feature channel to the RadioUNet, 
real-life measurements would be given to the RadioUNet in the real-time operations, 
even though measurements from the crude simulation are used in training.
Real-life measurements can be provided in real-time directly from the deployed devices, e.g., from the beacon signals of the transmitters, in the same way current systems report ``Channel Quality Indicators'' as measurements of the received signal strength.  Hence,  no costly measurement campaign is needed for training. 
The network can generalize well to real-life radio maps since it learned to interpolate the measurements, 
which are now accurate, while what was learned from the crude simulations roughly guides the interpolation procedure to be physically feasible. We demonstrate this experimentally by training on coarse simulations and using IRT4 samples and targets (as a proxy for real-life measurements) in testing.}

\par{\noindent\textbf{Large and dense simulation dataset $+$ small sparse measured dataset.}} 
Here, in addition to the large dataset of dense measurements, \textcolor{black}{we also assume that we have a small dataset of sparse measurements taken from refined radio maps (the IRT4 simulations, which can be potentially replaced by real-life measurements).} 
For each of the 700 maps of the RadioMapSeer dataset we consider two transmitter locations, and measurements in $K$ receiver locations, where $K$ is fixed, e.g., $K=300$. %The measurements are taken from the IRT4 dataset, which serves a surrogate to real life measurements.
In this scenario we first train a large network that estimates the crude simulations, using the large simulation dataset. Then, we improve the network output, using a smaller network, to match the small dataset of real-life measurements (see Subsection \ref{Adaptation to Measurements}). %This procedure reduces the error by a factor of 1.5 to 2.

%\subsubsection{Simulation scenarios}
%\label{Simulation scenarios} $ $
%
%
%\par{\noindent\textbf{Accurate simulation.}}
%In the first scenario we assume that the simulated radio maps from the dataset are the ground-truth, representing real %life radio maps.
%\vspace{1mm}
%
%\par{\noindent\textbf{Unreliable simulation.}}
%In the second scenario we assume that the simulated radio maps are only approximations to the ground-truth physical %phenomenon. This scenario is realistic, since IRT2 and DPM are rough approximations that capture  the correct shadow %pattern of radio maps, but lack fine details. 
%
%We can model this idea in two ways. First, we train the network on one type of simulation, but then test it on another %type of simulation, assumed to represent real-life. Second, we can produce a non-deterministic dataset, in which the radio %map due to a given city map and transmitter location is some weighted average of the two types of simulations, DPM and %IRT2, with random weights. %No network can fully learn the physical phenomenon due to the randomness, and thus needs to %rely on measurements.
%
%The goal in the randomized simulation is to demonstrate to the network that it can only rely on the simulations for the %big-picture behavior of radio maps, shared both by DPM and IRT2, but not for the fine details. This pushes the network to %use additional information for refining the estimations, like the input measurements if given, or the smaller dataset of %sparse measurements if given.

%%%%%%%%%%%%%%%%%%%%%%%%%%%%%%%%%%%%%%%%%%%%%%%%%%%%%%%%%%%%%%%%%%%%%%%%%%%%%  HERE HERE HERE 
\subsection{RadioUNet Architectures}

\begin{figure}
	\centering
		\includegraphics[width=1\linewidth]{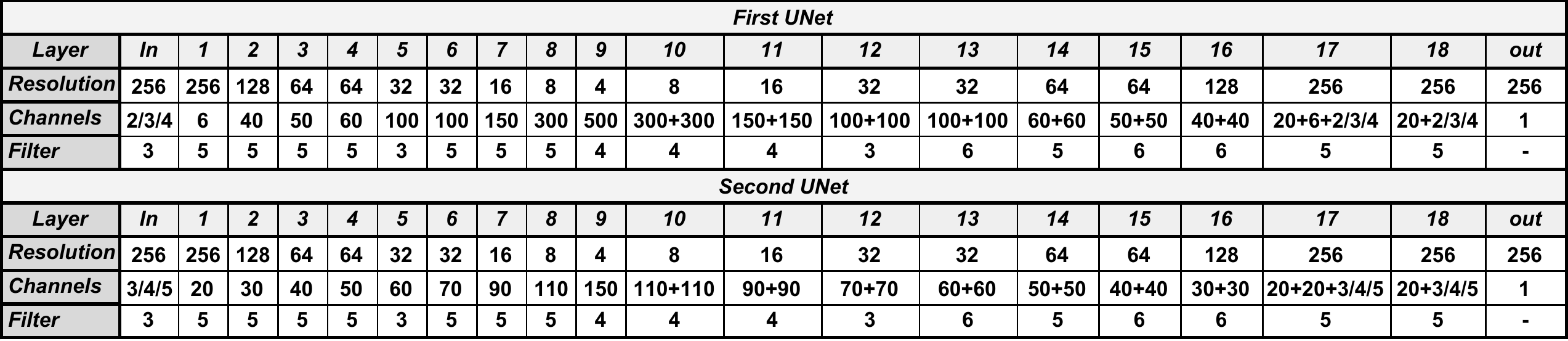}
     \caption{\small RadioUNet architecture. \emph{Resolution} is the number of pixels of the image in each feature channel along the $x,y$ axis. \emph{Filter} is the number of pixels of each filter kernel along the $x,y$ axis.
     The input layer is concatenated in the last two layers.}
     \label{tabele1}
\end{figure}

The simplest RadioUNet comprises of one UNet. The input of the UNet has two, three or four feature channels, depending if measurements and cars are used, and the output is the one feature channel estimated radio map.
In most architectures of RadioUNet we compose a second UNet on the first one. 
We call such an architecture a WNet (U$+$U makes a W). 
The input of the second UNet are the same as the inputs of the first UNet, plus an additional feature channel, the output of the first UNet. The architectures of our proposed UNets are reported in Table \ref{tabele1}. \textcolor{black}{The number of layers and feature channels were crudely searched to reduce overfitting and increase performance on the validation set, while not being too large to allow fast inference.}
The second UNet can be used for three different purposes, summarized in the following three subsections.

\subsubsection{Retrospective Improvement}
The idea here is to give RadioUNet a chance to improve its estimation in retrospective. The first UNet learns 
implicitly an algorithm for estimating the radio map from the input, by extracting high level concepts from the map and synthesizing a radio map from them. The philosophy here is that it would be beneficial to inspect the resulting estimation, and correct visible inconsistencies with the map and with the physical phenomenon.
To inspect the output of the first UNet, a second UNet extracts high level concepts from the estimated 
radio map, the city map, \textcolor{black}{and all other inputs,} and synthesizes from these concepts an improved estimation of the radio map.
We observe that the retrospective improvement yields better performance especially when the first UNet is small (see 
Fig.~\ref{fig:WNET_sizesA}). This WNet is thus a technique for reducing the size of the RadioUNet without degrading performance.

The WNet is trained in a curriculum. The first UNet is trained first to estimate the ground truth radio maps, with MSE loss. In the second phase, the weights of the first UNet are frozen, and the second UNet is trained to estimate the ground truth radio maps with MSE loss.

\subsubsection{\textcolor{black}{Adaptation to Refined Measurements}} 
\label{Adaptation to Measurements} 
Here, we first train the first UNet to estimate coarse simulations from the large dataset with MSE loss. The simulations may be randomized or deterministic. After training, the weights of the first UNet are frozen, and the second UNet is trained to improve the estimation of the first UNet on the small dataset of IRT4.

The IRT4 training consists of sparse images, namely, for each map, there are $K$ Rx locations $\{x_k\}_{k=1}^K$, and the pathloss $f(x_k)$ is only known for these locations. We typically take $K=300$.
The loss function for the second UNet is the weighted MSE, with weights $W_k=\frac{1}{K}$ for the points $\{x_k\}_{k=1}^K$, and weight 0 for the unmeasured points. We train the adaptation UNet in two steps. First, we train a retrospective improvement UNet on the coarse dataset, and then we further train this UNet on the sparse IRT4 dataset.

\subsubsection{Thresholder} 
A thresholder second UNet is used in the service area classification method. The goal of the second UNet here is to take the estimated radio map of the first UNet and to produce a service map from it. More details are give in Subsection \ref{Service area classification}.

\subsection{Training}

\begin{figure}
	\centering
	\includegraphics[width=0.99\linewidth]{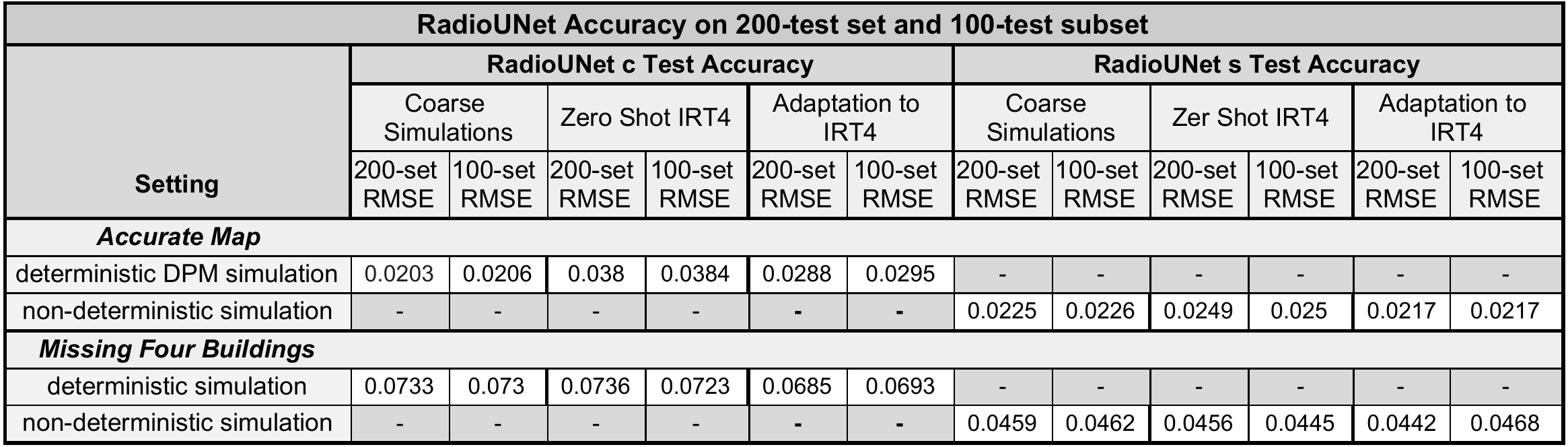}
	\caption{\small \textcolor{black}{Performance of selected RadioUNet methods on the 400/100/200 train/validation/test split. The last 100 test examples of the 200 test set are the 100 test examples of the original 500/100/100 split. The performance on the 200 and 100 test sets is comparable, indicating that the 100 test set is not ``too small''.}}
	\label{tabele_split}
	\end{figure}

The 700 maps of the RadioMapSeer dataset are \textcolor{black}{randomly} split into 500 training maps, 100 validation maps, and 100 test maps. \textcolor{black}{The random split is fixed, and available in the project web page\footnote{\url{https://github.com/RonLevie/RadioUNet}}.}
\textcolor{black}{We aim in this split to have a large enough training set, for avoiding overfitting, and a large enough test set, to avoid test set bias. To illustrate that this split is reasonable, we also consider a 400/100/200 train/validation/test split, where the last 100 test example of the 200 are the 100 test examples of the original 500/100/100 split. After training on the 400/100 train/validation split, the error on the 200 test set is very close to the error on the last 100 test examples (see Fig.~\ref{tabele_split}). Hence, there does not seem to be a visible bias in the original 100 test set.}

We perform supervised learning on the RadioMapSeer dataset.
The loss function is the MSE between the inferred radio maps by RadioUNet and the simulation radio maps from the training set. Training of all methods was performed with Adam \cite{Adam}, with learning rate of $10^{-4}$. We take 50 epochs for each UNet, no regularization, and batch size 15. To alleviate overfitting, out of the 50 epochs we pick the model with smallest error in the validation set. Lastly, the models are tested either on the coarse simulations on the test maps, or on the IRT4 simulations on the test maps. Performance is evaluated by RMSE on the gray levels and by NMSE (normalized MSE). Note that the RMSE in dB is 80 times the RMSE of gray level.

%For low intensity noise floors we train in curriculum learning.
%The idea is that it is easier to estimate only the high gains, since they roughly represent the first interactions in ray tracing. Lower gains represent higher order interactions. If UNet is trained directly to estimate the low noise floor truncated radio map, it learns a very smooth behavior, and misses the fine shadowing details.
%In the curriculum, the noise floor of the target radio maps is initially high and gradually decreases until reaching the desired level.

\subsection{RadioUNet Performance}

%\subsection{Transferability of the RadioUNet}

\begin{figure}
	\centering
	\includegraphics[width=0.99\linewidth]{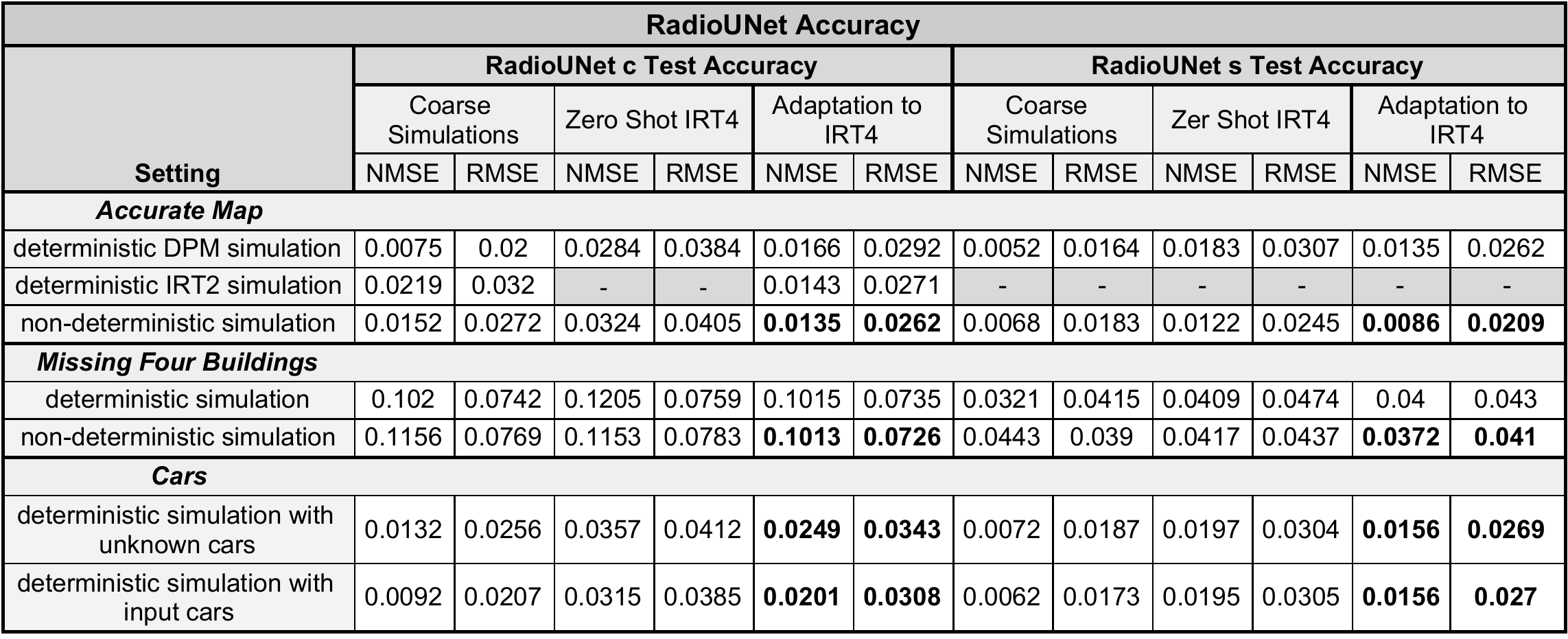}
	\caption{\small Comparison of RadioUNet accuracy in different scenarios}
	\label{tabele2}
	\end{figure}

In Fig.~\ref{tabele2} we report the results in all of the above settings. Recall that RadioUNet$_{\rm C}$ and RadioUNet$_{\rm S}$ denote the RadioUNet based on no input measurements and input measurements, respectively.
From the table we can observe that both the adaptation method to sparse IRT4 samples, and the training with randomized coarsely simulated maps, promote transferability.
%
%both transferability strategies are successful -- both the adaptation of UNet to sparse IRT4 and the non-deterministic coarse simulations. 
%\textcolor{black}{This last sentence is very unclear to me .. do you mean: ``both the training of UNet with sparse ITR4 samples, and the training with randomized coarsely simulated maps ????}
%
%
%
%
All accuracies are given both in NMSE and RMSE. 
     RMSE is the square root of the MSE on the whole test set.  
     The pathloss threshold is taken as $\textup{P}_{\textup{L,thr}} = -127$dB. %RMSE in dB is 80$*$RMSE.
     The best results on IRT4 for each category are marked in bold face.
     RadioUNet$_S$ was trained and tested with a random number of input measurements between 1 and 300.
     \emph{Zero-shot IRT4} means testing the methods, trained on coarse simulations, on IRT4. \emph{Adaptation to IRT4} means training a second small UNet to match the sparse IRT4 measurements. 
     All architectures are based on the WNets of Fig.~\ref{tabele1}, where for zero-shot transfer the second UNet is a retrospective improvement, and for adaptation to sparse IRT4, the second UNet is the adaptor.
     The receiver points of the sparse IRT4 dataset are randomly generated for each map, and fixed forever.
     For RadioUNet$_C$, the sparse IRT4 dataset has 300 receivers per transmitter. For RadioUNet$_S$, the sparse IRT4 dataset has 600 receivers per transmitter, out of them 1 to 300 random points are taken as input points of the RadioUNet$_S$. The training loss is computed for all 600 points. To show that the higher transferability of the random simulations is not simply because IRT2 is closer to IRT4 than DPM, we also include  the scenario where the deterministic simulation is IRT2. This produces inferior results compared to the random simulations.

%We consider a random number of measurements, between 10 and 300.
%
%RadioUNet estimates the radio maps from the coarse simulation test set very accurately and efficiently. However, the RadioMapSeer dataset consists of crude simulations of radio maps, while the end goal is to predict radio maps in real life. It is thus important to show that RadioUNet, trained on simulations, performs well in real life. 
%As explained in the introduction, the fine simulations IRT4 from the RadioMapSeer dataset serve as  surrogates for real life pathloss functions. 
%In Table ** we compare the performance of the proposed transferability approaches.
%The transferability results are reported in Table***** in Appendix ****.

%To show that our networks are robust, we also test transferability between different settings on RadioUNet.
%Namely, we train RadioUNet in one setting, and test it in another. The results are reported in Table ** in Appendix ****.

%Aside from the above transferability approaches, we can also roughly bound the performance in real-life of the RadioUNet, trained only on the DPM dataset with no adaptation to sparse realf-life measurements.
%DPM is roughly estimated to have accuracy of order $10^{-2}$ NMSE \cite{dominantPathUrban}, which is orders of magnitude higher than the error between the RadioUNet and DPM, an order of $10^{-2}$ NMSE. Thus, by the triangle inequality, the estimated error between RadioUNet and real life radio is dominated by an order of $10^{-2}$ NMSE. Note that this only roughly bounds the RMSE by an order of $10^{-1}$.

\begin{figure}
\centering
\begin{subfigure}{0.60\textwidth}\centering
\includegraphics[width=0.61\linewidth]{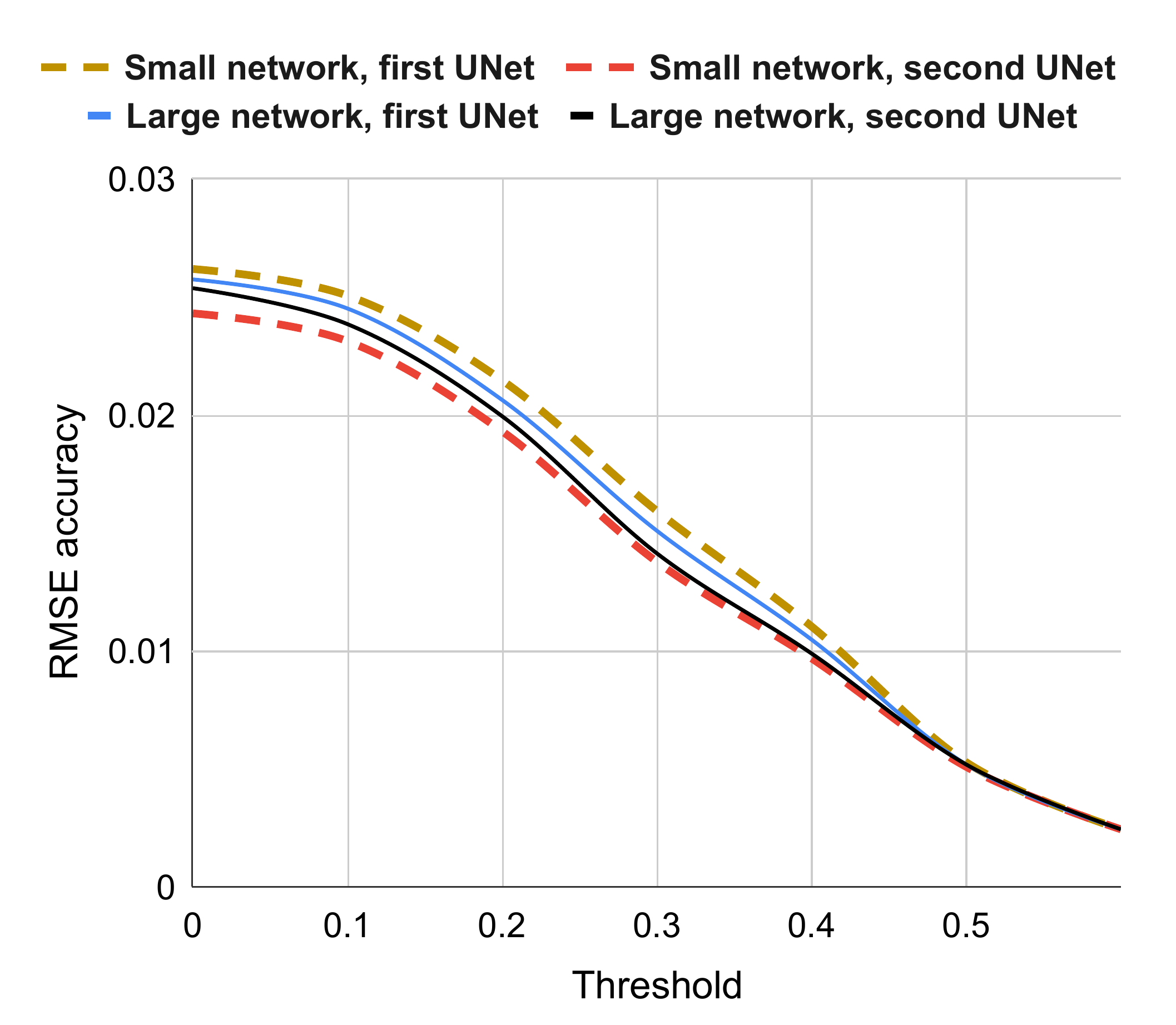}
 \caption{\small Accuracy of RadioUNet$_C$ of different network sizes and different pathloss thresholds $\textup{P}_{\textup{L,thr}}$. 
    The small network has 6,109,271 parameters and large one has 25,411,831 parameters.
    We plot the accuracy of the RadioUNets with and without the retrospective improvement. Small networks outperform large networks when both have retrospective improvement. The accuracy of RadioUNet with pathloss threshold at pixel value 0.6 is comparable to the quantization error of the png image file. }
    \label{fig:WNET_sizesA}
\end{subfigure} $\ $
\begin{subfigure}{0.35\textwidth}\centering
\includegraphics[width=0.99\linewidth]{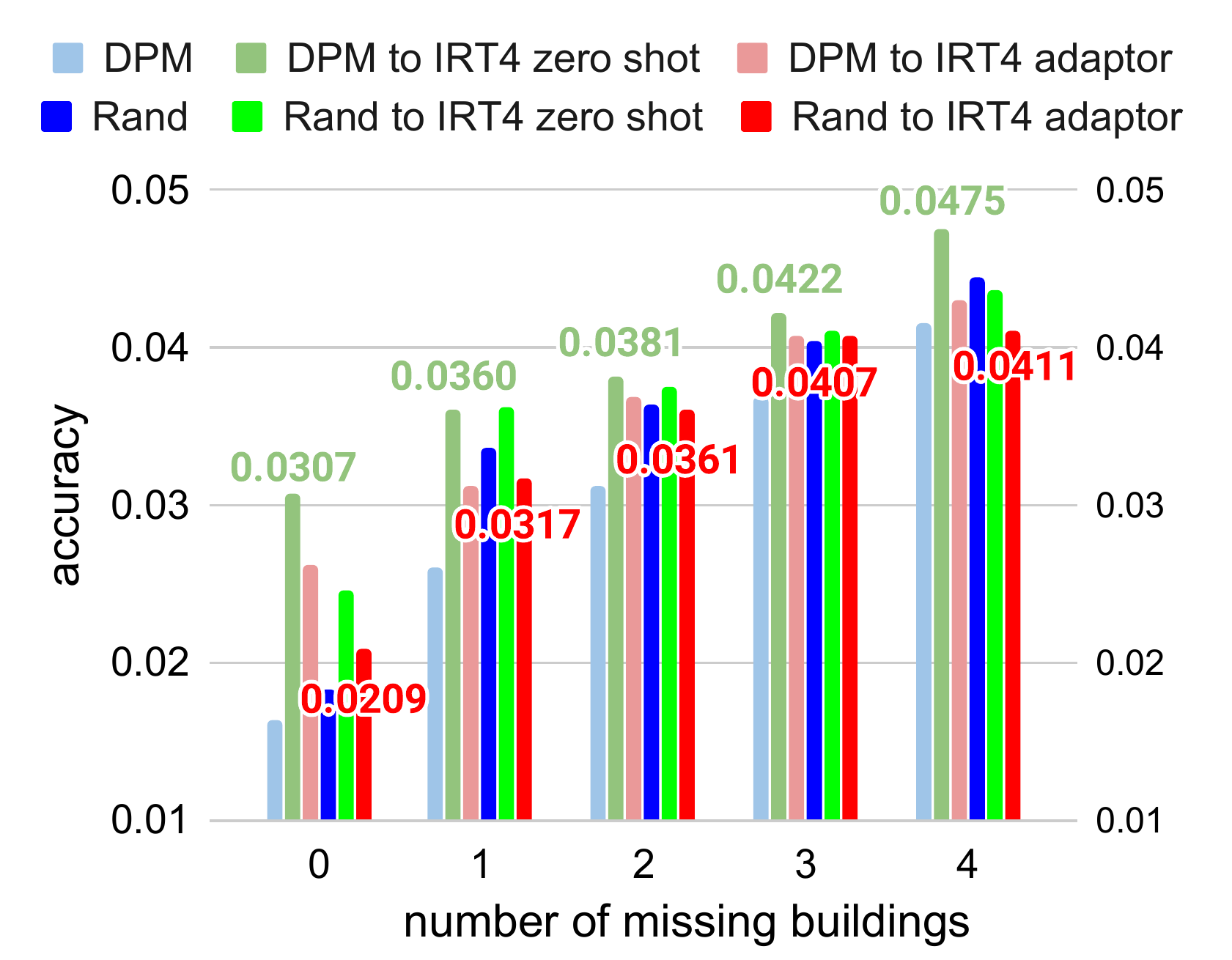}
		 \caption{\small  Accuracy of RadioUNet$_S$ with different numbers of missing buildings, different types coarse simulations, and different transfer methods to sparse IRT4. }
    \label{fig:WNET_sizesB}
\end{subfigure}		
\caption{RadioUNet performance}		
		\end{figure}

In Fig.~\ref{fig:WNET_sizesA}  we compare RadioUNet$_{\rm C}$ with and without retrospective improvement for different pathloss thresholds. The results demonstrate that the retrospective improvement UNet is effective when the first UNet is small, thus making it a useful strategy for reducing the network size for the same accuracy. 
In Fig.~\ref{fig:WNET_sizesB} we compare the performance of different RadioUNet$_{\rm S}$ methods on maps with various numbers of missing buildings. We observe that the strategy of combining random coarse simulations with an adaptor UNet to IRT4 promotes transferability. %The effectiveness of the method decreases as the number of missing buildings increases. 
%\textcolor{black}{I think that this last sentence is obvious ... I mean, it is obvious that as you perturb the maps more and more the accuracy decreases .. it would be a phenomenal surprise the opposite .... so, I wonder if this redundant sentence is even necessary since it seems to ``demote'' the effectiveness of the approach. Also, do we have alternatives? for example, if there is an alternative approach for which the accuracy does not decrease as you keep adding more and more perturbations it would make sense .. but in the absence of such term of comparison, .. we do not know how effective the method is since we do not know a fundamental limit ... (e.g., an information theoretic limit, which here is of course out of reach ... this means that maybe the method is efficient because it does the best possible, but simply the fundamental limit of the possible accuracy says that as you increase uncertainty the achievable accuracy decreases ...}

%\begin{figure}
%\centering
%		\includegraphics[width=0.40\linewidth]{Accuracy of RadioUNet sizes3.pdf} $\quad$
%		\includegraphics[width=0.40\linewidth]{MissBuild15.pdf}
%		 \caption{\small \textbf{Left}: accuracy of RadioUNet$_C$ of difference network sizes and different pathloss thresholds $\textup{P}_{\textup{L,thr}}$. 
%    The small network has 6,109,271 parameters and large has 25,411,831 parameters.
%    We plot the accuracy of the RadioUNets with and without the retrospective improvement. Small networks outperform large networks when both have retrospective improvement. The accuracy of RadioUNet with pathloss thresholds at pixel balue 0.6 is comparable to the quantization error of the png image file. \textbf{Right}: Accuracy of RadioUNet$_S$ with different numbers of missing buildings, different types coarse simulations, and different transfer methods to sparse IRT4. }
%    \label{fig:WNET_sizes}
%		\end{figure}

%%%%%%%%%%%%%%%%%%%%%%%%%%%%%%%%%%%%%%%%%%%%%%%%%%%%%%%%%%%%%%%%%%%%%%%%%%
\section{Comparison of RadioUNet to state-of-the-art}

In Fig.~\ref{fig:Second_figure} we present the performance of different methods of radio map estimation. For methods that depend on samples, we use an input map with four missing buildings, and for methods that do not rely on samples we use the full map.
Apart from the fact the RadioUNet outperforms the data driven interpolation methods, the tomography method and the previously proposed deep learning approach significantly, these other methods need a separate training/optimization to fit the model to {\em each} map. Particularly,  variations  in  the  environment, like moving cars,  requires re-computing the methods, which is not efficient. RadioUNets, in comparison, are trained offline only once, and are then employed in any environment very efficiently. 
 \textcolor{black}{RadioUNet can deal with cars by using the measurements input, where the network is trained on a dataset of simulations with cars.}
All GPU methods ran on Nvidia Quadro GP100, and CPU methods on Intel Core i7-8750H.

\subsection{Comparison to Model-Based Simulation}
We compare the run-time\footnote{Notice that the run-time is the computation time of the {\em trained} network. This does not include the training, which is done offline and once for all.} 
with the efficient dominant pathloss method \cite{dominantPathUrban}. RadioUNet estimates radio maps 
roughy two to three orders of magnitudes faster. In our experiments,
WinProp completes a simulation in roughly an order of $1$sec on a Intel Core i7-8750H CPU, and RadioUNet an order of $10^{-3}$sec to $10^{-2}$sec. % In addition, RadioUNet has the capability to adopt 
  IRT2 and IRT4 took an order of $10$sec and $10^{2}$sec, respectively.

\subsection{Comparison to Data Driven Interpolation}

Next, we compare RadioUNet$_{\rm C}$ and RadioUNet$_{\rm S}$ with data driven interpolation methods: radial basis function (RBF) interpolation using multiquadric function \cite[Sect 5.1]{BishopNNPR} and tensor completion \cite{TensorCompletion}. For the data driven methods we set to zero the gray level values inside the known buildings of the map post-processing, thus using the urban geometry data. Without this step, data driven interpolation methods obtain a very poor accuracy since they are not able to recover the sharp building edges.
 In Fig.~\ref{fig:Second_figure}, we plot the average NMSE over 80 Txs of RadioUNet$_{\rm S}$ and of the two data driven interpolation methods as a function of the number of samples. Both versions of RadioUNet clearly outperform state-of-the-art. Aside from that, RadioUNet is roughly three orders of magnitude faster than RBF interpolation, and five orders of magnitude faster than tensor completion interpolation.

%\textcolor{black}{Say that in the interpolation papers they used an unrealistic number of samples, taken as some percentage of all pixels. e.g., 10. This would mean 6553 in our situation.}

\subsection{Comparison to Model-Based Data Fitting}

We compare RadioUNet with a tomography method.
In general, tomography methods model the attenuation in the channel strength as the sum of a distance dependent pathloss and a shadowing term which models the attenuation due to obstructions. To model shadowing, a spatial loss field $L:\mathbb{R}^2\rightarrow \mathbb{R}$ (SLF) is defined. For each spatial location $y$, the value $L(y)$ in a sense models the transparency of $y$, where $L(y)=0$ models free space, and $L(y)>0$ represents a ``translucent'' obstacle. The shadowing term from the Tx location $x$ to the Rx location $y$ is computed as the integral of $L$ in a narrow oval for which the transmitter and receiver sit on the edges of he largest diameter. More generally, the oval can be replaced by some other shape, which may be trainable.

Note that as opposed to ray-tracing methods, tomography method do not consider at all wave propagation phenomena like diffraction and reflections, and only model the attenuation due to the penetration of the signal through material. For high frequency signals, the attenuation due to penetration in urban environments is very large, which make tomography method less realistic than DPM and IRT.

In tomography methods (e.g., \cite{BlindTomo,ElasticNets,LowRankOutliers,AdaptiveBayesian,VariationalBayes}), the SLF is typically estimated from observed pathloss values between samples transmitter-receiver pairs, by solving an inverse problem.  In our situation the problem is easier, since we are given the city map. Thus, the SLF outside the buildings, in free space, is known to be zero.
Moreover, the building material is constant, and thus it is natural to consider an SLF with one value $f$ inside buildings, and $0$ outside. Hence, the computation of the SLF is reduced to finding the scalar $f$ for which the tomography method gives a radio map as close as possible to the ground truth radio map. This method takes an order of $10^2$sec to run. %\textcolor{black}{As the weighting function we use the \emph{normalized ellipse function} \cite{TomoProp} and discard the possibility of learning the weighting functions \cite{BlindTomo,ElasticNets}, which require training.}

%As pointed out in \cite{V2VOnlinePL}, tomography methods consider fixed environments for real-time radio map estimation. Any change in the environment requires a new training.

%\textcolor{black}{We need to mention the weighting function, as we use heuristics for weighting functions when we compare them with us. Mention blind methods \cite{BlindTomo,ElasticNets} here. Notice that with the learned weighting functions, the tomography method, which with given SLF  we compare our clean results, will perform much better. However, learning the weights would add data collection and computation delays. Therefore, we only compare with fixed weighting functions.}

\subsection{Comparison to Deep Learning Data Fitting}

We compare RadioUNet to the deep learning one-step prediction approach of \cite{DocomoTwoStep}. We note that the two-step prediction approach of \cite{DocomoTwoStep} did not perform well in our setting.
As explained in Subsection \ref{Radio Map Prediction Using Deep Learning}, this method is a data-fitting of a multilayer perceptron (MLP) to a 4D radio map of a specific city map. The network receives the transmitter and receiver 2D locations and returns the estimation of the pathloss for this pair. The network architecture is reported in Fig.~\ref{fig:Second05_figure}. For a fixed map, the 80 transmitters are split to 60 training, 10 validation and 10 test transmitters. The network is trained and tested against all receiver locations in the 
256$\times$256 grid. This method takes an order of $10$sec to estimate all 256$\times$256 pixels, which must be computed separately.

\subsection{\textcolor{black}{Complexity comparison}}
Consider a radio map of $n\times n$ pixels. We compare the asymptotic complexity of all considered methods.
\begin{itemize}
    \item 
    \emph{RadioUNet.}
    In our UNets, the resolution of Layer $l$ is $O(n^24^{-l})$, and the number of channels $c_l$ increase in $l$ slower than $2^l$. Moreover, the convolution kernels are spatially localized. This means that the complexity is $O\big(n^2\sum_l 4^{-l}c_lc_{l-1}\big)=O(n^2)$.
    %Since the number of layers in a UNet is of order $\log(n)$, and the convolution kernels are localized, the complexity of RadioUNet is $O(n^2\log(n))$ for images of $n\times n$ grid points.
   \item
    \emph{MLP regression.} Each estimation takes $O(1)$ operations, and there are $n^2$ estimations to cover the map. Thus the complexity of $O(n^2)$. We note that in practice the constant here is large, since the MLP must be able to realize a complicated function on a 4D domain.
    \item
    \emph{DPM and IRT.} In both methods there is a pre-processing step in which a graph that represents the line of sight between different wall and edge segments, and receiving points, is constructed. For a map with $W$ wall and edge segments. The complexity of this step is $O(W(W+n^2))$. Let us roughly estimate this complexity in terms of $n$. Since walls are one dimensional, a reasonable estimation of the number of wall pixels in a dense urban environment is $An$ for some $A>1$. Another reasonable assumption is that, on average, each set of $B>1$ wall/edge pixels are grouped to one segment. This puts the pre-processing complexity at $O(n^3)$. The pre-processing complexity poses a lower bound to the complexity of both DPM and IRT, regardless of the number of interactions. In IRT, the complexity grows exponentially with respect to the number of interactions (multiplied by $n^2$). In another version of DPM, preprocessing is not required \cite{Wolfle_fieldstrength}.
    In this algorithm, the so called \emph{step 3} dominates complexity, with $O(n^3)$ operations. Indeed, the average path has $O(n)$ pixels, and there are $n^2$ receiving points.
    %\textcolor{black}{C: I've gone through the documentation of the software. There is apparently no preprocessing for the DPM. However, I couldn't find a paper on DPM yet (seems like there is none) that explains how the dominant path is chosen. In the documentation, there is no visibility tree construction anymore (as in the paper on urban DPM that I sent you) and no explanation how the dominant path is found. For IRT2, it turns out I used the default "accelerated" preprocessing, which only considers one interaction (reflection or diffraction) of the tiles/segments. The complexity is then $O(W(n^2))$, instead of $O(W(W+n^2))$. There is not much difference between the preprocessing times of these two options ($W << n^2$): Single interaction: 34secs, multi interactions: 37secs. For IRT4, multi interactions is necessary in the preprocessing. Computation times after preprocessing: IRT2: 8sec, IRT4:17sec. These results are from a map with 52 buildings. Currently, all we can say is that DPM requires no preprocessing and has the lowest computation time.}
    \item
    \emph{Tomography.}  On average, the shape in which the spatial loss field is integrated has proportional area to $n^2$. Thus, since all receiver points are computed separately, the complexity is $O(n^4)$ with a small constant.
    \item
    \emph{RBF.}
    For $k$ measurements, RBF takes $O(k^3)$ operations. In typical situations we take $k$ proportional to $n$, in which case the complexity is $O(n^3)$.
    \item
    \emph{Tensor completion.} In our case we have a matrix completion problem. 
    The dominating term in each iteration is due to SVD within the shrinkage operation \cite{Gandy_2011}. Thus the complexity for $T$ iterations is $O(Tn^3)$. 
    %In our case the problem is actually a matrix completion problem. The paper we cite adds a regularization term (we considered the TV2 variant, which they reported to be the best in small sample regime) to the paper:" S. Gandy, B. Recht, and I. Yamada, “Tensor completion and lown-rank tensor recovery via convex optimization,". In this work it is argued that dominating complexity term is due to SVD within the shrinkage operation. Complexity of SVD for an  $n\times n$ matrix is $O(n^3)$. The proximal map of the TV2 regularization consists of inversion of a tridiagonal Toeplitz matrix which is supposed to be solved efficiently (as referenced in TC paper we cite). Futhermore, these operations are repeated in sequel until the maximum number of iterations is reached (I used 1000 as max)
    %\textcolor{black}{(Ron) so it is $O(n^3)$ times the number of iterations?
    %}
\end{itemize}

%%%%%%%%%%%%%%%%%%%%%%%%%%%%%%%%%%%%%%%%%%%%%%%%%%%%%%%%%%%%%%%%%%%%%%%%%%%%%%%%%%%%%
\section{Applications}

In this section we demonstrate the usefulness of RadioUNet with two simple applications and also discuss some future applications as future work.

\subsection{Coverage Classification}
\label{Service area classification}

Service area classification shows up in two situations. In the first problem, given a Tx-Rx link, we would like to know if the received signal strength is large enough. 
In the second problem, given two Tx-Rx links, we would like to know if the interference caused by one link on the other is low enough. In both cases, the goal is to classify if the pathloss of a certain Tx is above or below some threshold at the location of some Rx.  
%\textcolor{black}{PLEASE CHECK THE FOLLOWING IN THE ORIGINAL VERSION THERE WAS A TYPO AND I WANT TO MAKE IT 
%CONSISTENT WITH THE REST OF THE PAPER ...}
\textcolor{black}{
For a fixed Tx location $x$, let $f(y)$ denote the radio map at location $y$. 
We define the \emph{coverage map} as the thresholding function 
\begin{equation} 
C(y) =
\begin{cases}
0 &  \text{if } f(y) \leq T, \\
1   & \text{if } f(y) > T, \\
\end{cases} 
\end{equation}
where $T$ is a threshold in gray scale.}  
For the first problem, depending on the system requirements, $T$ is some value above the noise floor. 
For example, for high bit rates the signal has to arrive with high SNR, so a typical value for $T$ might be pixel value 0.5 (see e.g. \cite{etsi_302663}). For the second problem, a typical choice for $T$ is the noise floor, which is pixel value 0.2 for us. 

Our goal is to predict the coverage map from the input city and transmitter location.
%On the one hand, the coverage map is a function of the radio map, and thus does not contain more information than the radio map itself. On the other hand, the coverage map encodes less information than the radio map, since all gray levels are converted to 0 or 1. 
%
Note that in principle UNets are expressive enough to predict coverage maps, since coverage maps are a sub-phenomenon of radio maps, and UNets are expressive enough to predict radio maps. However, this naive point of view disregards 
the fact the the gradient descent optimization procedure is highly non-exhaustive, and only searches parameter configurations along a 1D path.
As it turns out, simple UNets fail to learn meaningful predictions of coverage maps.
Intuitively, %the shadow patterns of 
 radio maps are more predictable than coverage maps since shadow patterns are always associated with simple concepts like building corners and spatial relations between building, receiver locations, and the location of the transmitter. 
In contrast, in the coverage map most shadow edges disappear and are ``absorbed'' by one or the domains 
above or below $T$.

For the architecture to successfully predict the coverage map, it must first understand the underlying phenomenon of radio maps. We thus consider a WNet architecture, where the first UNet  is RadioUNet, and predicts the radio map from the city and transmitter inputs, and the second UNet receives the predicted radio map as input, along with the map and the transmitter location, and computes the coverage map from them. We call the second UNet the thresholding UNet, or TUNet. We call this architecture the Coverage WNet, or CWNet in short.

To train CWNet we use curriculum learning. We first train the RadioUNet as before. We then freeze the RadioUNet, and train the TUNet in a curriculum as explained next. 
As it turns out, the discontinuous nature of the coverage map is still too challenging for the TUNet to learn directly. Instead, we relax the coverage map to a soft coverage map
$C_\alpha(y) = {\rm sigmoid}\Big(\alpha\big(f(y) - T\big)\Big)$
where $\alpha$ is a parameter that determines how soft the transition between 0 and one 1 is. 
We interpret $C_\alpha(y)$ as the probability of location $y$ being in the coverage area.
In the curriculum we first train the TUNet to predict $C_\alpha(y)$ with $\alpha=1$, and gradually increase 
$\alpha$. We end up with $\alpha=128$, which we judge to be high enough to represent a sharp transition.

\begin{figure}[!ht]
	\centering
	\begin{subfigure}{0.25\textwidth}\centering
\includegraphics[width=1\linewidth]{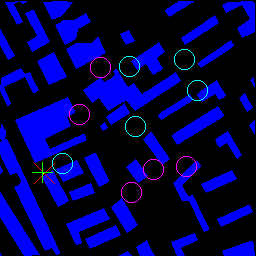}
\caption{\small Localization}
\label{fig:local}
	\end{subfigure}		
	\begin{subfigure}{0.27\textwidth}\centering
		\includegraphics[width=0.93\linewidth]{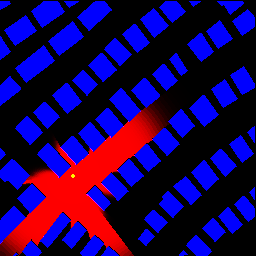}
		\caption{\small Ground truth coverage map}
	\end{subfigure}%
	\hfill
	\begin{subfigure}{0.25\textwidth}\centering
		\includegraphics[width=1\linewidth]{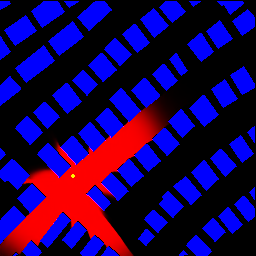}
		\caption{\small Estimated coverage map}
	\end{subfigure}%
	\hfill
	%\begin{subfigure}{0.19\textwidth}\centering
	%	\includegraphics[width=1\linewidth]{Thr2L78target2.png}
	%\end{subfigure}%
	%\hfill
	%\begin{subfigure}{0.19\textwidth}\centering
	%	\includegraphics[width=1\linewidth]{Thr2L78predict2.png}
	%\end{subfigure}%
	\begin{subfigure}{0.21\textwidth}\centering
		\includegraphics[width=1\linewidth]{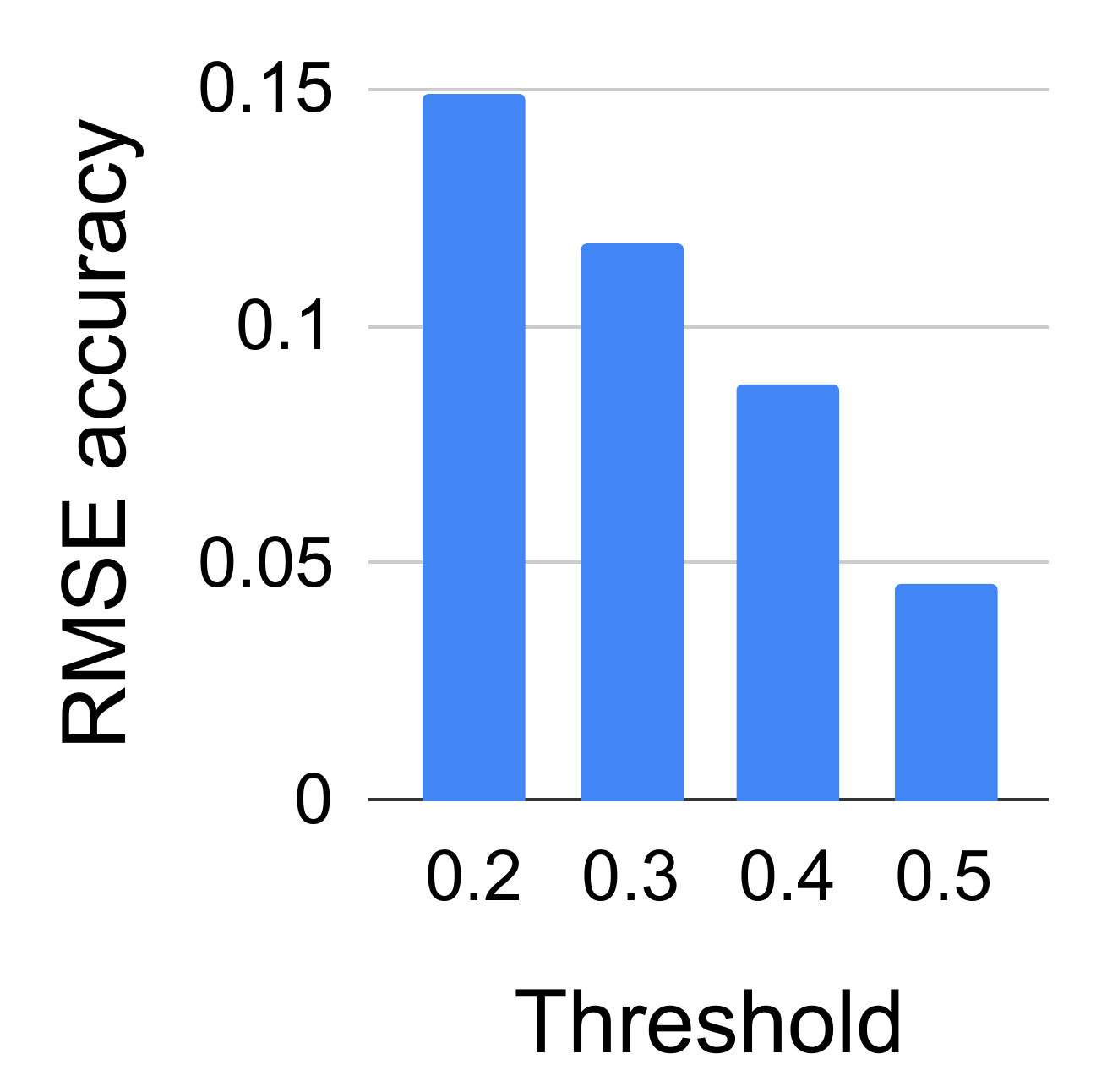}
		\caption{\small Accuracy of service map estimation}
	\end{subfigure}
	\caption{\small \textbf{Left}: Localization result. Green $\rm +$: true Rx position, red $\rm X$: estimated Rx position, yellow: pixels of the localization intersection, magenta circles: Txs of the best localization result out of the $R$, green circles: the rest of the Txs. \textbf{Middle}: Coverage map results with threshold 0.5. Red: coverage map. Blue: city map. Yellow: transmitter. \textbf{Right}: accuracy of service map estimation for different thresholds in RMSE.}
	\label{fig:SWNet_acc}
\end{figure}

The accuracy of SWNet for different thresholds and an example service map are presented in Fig.~\ref{fig:SWNet_acc}.
%SerivceAccuracy
%\begin{figure}[!ht]
%    \centering
%    \includegraphics[width=0.3\linewidth]{SerivceAccuracy.pdf}
%    \caption{Accuracy of RadioUNets of difference sizes: .......}
%    \label{fig:SWNet_acc}
%\end{figure}

%The underlying phenomenon of pathloss is obscured in the morphological service area map. Even if a network has enough expressive capacity to represent the physical phenomenon, it will not explore the parameter configurations reguired for representing the physical phenomenon in a naive end-to-end loss function. Therefore, a more guided approach to steer the network at the correct direction is needed. We achieve this via a WNet architecture together with curriculum learning. 

%Superlevel set. Indicator function. Service map.
%To allow some uncertainty in service area classification, we consider a soft version of the service map. Given by
%\[{\rm sigmoid}\Big(D\big(R(x)-\mathcal{N}\big)\Big)\]
%for some large $D>0$. The larger $D$ is, the less soft the service map.

\subsection{Pathloss Based Fingerprint Localization}
\label{Pathloss Based Fingerprint Localization}

Suppose that a device is simultaneously in the coverage of several base stations located at Tx points $x_1, \ldots, x_K$, and reports the strengths $g_k$ (converted in gray scale) of their corresponding beacon signals. Let $f_k(y)$ denote the estimated radio map for Tx location $x = x_k$, for $k = 1, \ldots, K$. For some $\epsilon > 0$, we define the $\epsilon$-level set for level $g_k$ as
\begin{equation} \label{levelset}
L^\epsilon_k = \{ z \in \Gamma \; : \;  |f_k(z) - g_k| \leq \epsilon \},
\end{equation}
where $\Gamma$ is the discrete grid (domain of the radio map, in our case the 256$\times$256 grid). 
Then, in order to identify the location of the receiver $y$ we can consider the intersection of the $\epsilon$-level sets
$S = \bigcap_{k=1}^K L^\epsilon_k$. If this set is localized about a single point, then we have located $y$ %(withing the grid accuracy)
 with high probability. 

Assuming that the reported values $\{g_k\}_k$ are equal to the true radio map values, if for some $k$ the radio map prediction error satisfies $|f_k(y) - g_k| > \epsilon$ then $y \notin L^\epsilon_k$ and $y$ will not be contained in the intersection $S$. We call such $k$ an outlier.
In contrast, if we choose $K$ too small, then $S$ will contain multiple points and the localization is ambiguous. 
Hence, the method works well when the estimated radio maps are accurate and the number of reported signal strengths $K$ is large enough but not too large. 

%In practical cellular scenarios, a device can detect the beacon of some $K$ (which can be quite large when considering a urban environment with dense cell deployment) and  we shall select a subset of $K < N$ largest reported values. 
To alleviate the effect of outliers,  instead of computing a single intersection 
we can select random subsets of $J < K$ Txs and consider the intersection of the corresponding 
$\epsilon$-level sets. We also take random $\epsilon$ values for each map, since different maps have different unknown accuracies.  Repeating this random selection $R$ times, we generate $R$ candidate sets, some of which may be empty and some of which may contain multiple points. For the $R'$ non-empty outcomes 
we compute a score for the quality of the result, and pick the outcome with the best score.
For example, we use the variance of the localization outcome.
Let $S_t$ be the localization outcome of sample $t$, where $t=1,\ldots R'$. Then, 
we define the expected position given $S_t$ as
$\widehat{y}_t  = \sum_{z \in S_t} \frac{z}{|S_t|}$,
and the associated variance 
\[V_t = \sum_{z \in S_t}   \frac{|z-\widehat{y}_t|^2}{|S_t|},\]
where $|z - \widehat{y}|$ is the Euclidean distance between $z$ and $\widehat{y}_t$ in $\mathbb{R}^2$ and $|S_t|$ is the area of $S_t$. 
Since smaller variance means better localization, we pick the non-empty localization outcome 
with smallest variance.
%
%\textcolor{black}{(Ron) Explain is short how the pathloss localization can be combined with other information to improve localization.}
In this paper we mention this approach just as an example of the use of accurate radio map estimation. 
In future work we will deal with improving the pathloss based localization with more sophisticated 
localization extraction, and using additional signal fingerprints.

In Fig.~\ref{fig:local}  we present an example localization result with $K=10$, $J=5$, $R=5$, $\epsilon = 0.03$. The best outcome has a standard deviation of $0.5$ meters. %Green x-mark, red x-mark, magenta circles, cyan circles and yellow pixels visualize the true receiver position, estimated (center of mass) receiver position,  and $O_t$, respectively. 
The distance between the estimated and true receiver location is $1.58$ meters.

%\subsection{Additional Future Applications (need?)}
%
%Scheduling...?

\section{Conclusion}

In this paper we introduced RadioUNet, a deep learning method for simulating radio maps given a city geometry, Tx location, and optionally some pathloss measurements and car locations. For training RadioUNet, we introduced the new dataset RadioMapSeer, which we hope will be used for developing deep learning methods for pathloss prediction by other researchers as well. 
We developed approaches for transfering what was learned on the large dataset of coarsely simulated radio maps to real-life, and demonstrated the superior performance of our methods with respect to state-of-the-art, both in run-time and accuracy.

% conference papers do not normally have an appendix

% use section* for acknowledgment
%\section*{Acknowledgment}

%The authors would like to thank...

\subsection*{Acknowledgement }

The work presented in this paper was partially funded by the DFG Grant DFG SPP 1798 “Compressed Sensing in Information Processing” through Project Massive MIMO-II, and by the German Ministry for Education and Research as BIFOLD - Berlin Institute for the Foundations of Learning and Data (ref. 01IS18037A).

We thank Ibrahim Rashdan for a fruitful discussion on the impact of cars on the pathloss function.

% trigger a \newpage just before the given reference
% number - used to balance the columns on the last page
% adjust value as needed - may need to be readjusted if
% the document is modified later
%\IEEEtriggeratref{8}
% The "triggered" command can be changed if desired:
%\IEEEtriggercmd{\enlargethispage{-5in}}

% references section

% can use a bibliography generated by BibTeX as a .bbl file
% BibTeX documentation can be easily obtained at:
% http://mirror.ctan.org/biblio/bibtex/contrib/doc/
% The IEEEtran BibTeX style support page is at:
% http://www.michaelshell.org/tex/ieeetran/bibtex/
%\bibliographystyle{IEEEtran}
% argument is your BibTeX string definitions and bibliography database(s)
%\bibliography{IEEEabrv,../bib/paper}
%
% <OR> manually copy in the resultant .bbl file
% set second argument of \begin to the number of references
% (used to reserve space for the reference number labels box)

\bibliographystyle{ieeetr}
{\small
\bibliography{strings,pub2}
}

\end{document}